\documentclass[12pt]{JHEP3}
\usepackage{amsfonts}
\usepackage{amssymb}
\usepackage{lscape}
\usepackage{cite}
\usepackage{epsfig}
\usepackage{multirow}
\usepackage{longtable}
\usepackage[all]{xy}
\usepackage{amsmath}

\author{}

\newcommand{\drawsquare}[2]{\hbox{%
\rule{#2pt}{#1pt}\hskip-#2pt
\rule{#1pt}{#2pt}\hskip-#1pt
\rule[#1pt]{#1pt}{#2pt}}\rule[#1pt]{#2pt}{#2pt}\hskip-#2pt
\rule{#2pt}{#1pt}}
\newcommand{\fund}{\raisebox{-.5pt}{\drawsquare{6.5}{0.4}}}
\newcommand{\Ysymm}{\raisebox{-.5pt}{\drawsquare{6.5}{0.4}}\hskip-0.4pt%
         \raisebox{-.5pt}{\drawsquare{6.5}{0.4}}}
\newcommand{\Yasymm}{\raisebox{-3.5pt}{\drawsquare{6.5}{0.4}}\hskip-6.9pt%
        \raisebox{3pt}{\drawsquare{6.5}{0.4}}}

\newcommand{\be}{\begin{equation}}
\newcommand{\ee}{\end{equation}}
\newcommand{\ba}{\begin{array}}
\newcommand{\ea}{\end{array}}
\newcommand{\bea}{\begin{eqnarray}}
\newcommand{\eea}{\end{eqnarray}}

\newcommand{\ov}{\overline}
\def\IR{\relax{\rm I\kern-.18em R}}

\def\IP{\relax{\rm I\kern-.18em P}}
\def\inbar{\vrule height1.5ex width.4pt depth0pt}
\def\IC{\relax\,\hbox{$\inbar\kern-.3em{\rm C}$}}

\def\K3{{\bf K3}}

\def\ov{\overline}

\def\n2d{\cN_{V^*}^{\otimes 2}}

\def\IC{\mathbb{C}}

\def\IR{\mathbb{R}}

\def\IP{\mathbb{P}}

\def\cN{{\mathcal N}}

\def\nn{\nonumber}

\title{String Constraints on Discrete Symmetries in MSSM Type II Quivers}
\author{
Pascal Anastasopoulos$^{1}$\footnote{pascal@hep.itp.tuwien.ac.at},~
Mirjam Cveti\v c$^{2,3}$\footnote{cvetic@cvetic.hep.upenn.edu},~
Robert Richter$^{4}$\footnote{robert.richter@desy.de},~
Patrick K.S. Vaudrevange$^{5}$\footnote{patrick.vaudrevange@desy.de}~\\
$^1$ Technische Univ. Wien Inst. f\"ur Theoretische Physik, A-1040 Vienna, Austria\\
$^2$ Department of Physics and Astronomy,
University of Pennsylvania, Philadelphia, PA 19104-6396, USA\\
$^3$ Center for Applied Mathematics and Theoretical Physics,
University of Maribor, Maribor, Slovenia\\
$^4$ II. Institut f\"ur Theoretische Physik, Hamburg University, Germany\\
$^5$ Deutsches Elektronen-Synchrotron DESY, Notkestra{\ss}e 85, 22607 Hamburg, Germany\\
}
\date{}

\maketitle
\abstract{We study the presence of discrete gauge symmetries in D-brane semi-realistic compactifications. After establishing the constraints on the transformation behaviour of the chiral matter for the presence of a discrete gauge symmetry we perform a systematic search for discrete gauge symmetries within local semi-realistic D-brane realizations, based on four D-brane stacks, of the MSSM and the MSSM with three right-handed neutrinos. The systematic search reveals that Proton hexality, a discrete symmetry which ensures the absence of R-parity violating terms as well as the absence of dangerous dimension 5 proton decay operators, is only rarely realized. Moreover, none of the semi-realistic local D-brane configurations exhibit any family dependent discrete gauge symmetry.
}
\preprint{
DESY-12-182\\
TUW-12-32\\
UPR-1245-T\\
ZMP-HH/12-21}
 \clearpage
\begin{document}

\section{Introduction}

While the Minimal Supersymmetric Standard Model (MSSM) addresses various
short-comings of the Standard Model (SM), such as solving the hierarchy problem,
providing a natural dark matter candidate (i.e. the lightest supersymmetric
particle, LSP) and gauge coupling unification, it exhibits some severe
phenomenological problems, among them the issue of proton stability. The SM guarantees proton stability, whereas the MSSM allows
renormalizable R-parity breaking operators consistent with supersymmetry and
gauge invariance of the superpotential that do lead to a disastrous high proton decay rate.

The renormalizable SM gauge invariant superpotential terms read
\begin{align}
W_{MSSM} & = Y_U \,Q_L U_R H_u +  Y_D\, Q_L D_R H_d +  Y_L\, L E_R H_d + \mu\,H_u H_d \nn \\
         & + \lambda_1 \,U_R D_R D_R + \lambda_2\, Q_L L D_R  +  \lambda_3 \,L L E_R + \alpha \,L H_u \,\,,
\label{eq gauge invariant terms}
\end{align}
where the terms in the first line are the Yukawa couplings giving mass to
quarks and leptons after electroweak symmetry breaking as well as the
$\mu$-term. On the other hand, the second line contains terms that do not
conserve baryon and lepton number, so called R-parity violating terms. They can
lead to rapid proton decay, rendering the LSP unstable and thus eliminating the
possibility of any SUSY particle being the dark matter candidate. Moreover, SM
gauge invariance allows also for the dimension 5 proton decay operators 
\begin{align}
Q_{L} Q_L Q_L L \qquad \qquad U_R U_R D_R E_R\,\,,
\end{align}
which if not suppressed lead to a disastrous high proton decay rate.

Generically, discrete symmetries such as R-parity or Baryon triality are
invoked to forbid the presence of those superpotential terms.
Despite the fact that those discrete symmetries ensure the absence of such
undesired terms their origin remains unclear. There exist strong arguments
implying that in a consistent quantum gravity global symmetries, continuous or
discrete, are broken by quantum gravity corrections
\cite{Banks:1988yz,Abbott:1989jw,Coleman:1989zu,Kallosh:1995hi,Banks:2010zn}.
An exception are discrete symmetries that have a gauge symmetric origin, so called discrete gauge symmetries. For instance abelian discrete
symmetries $\mathbf{Z}_N$ are remnants of continuous $U(1)$ symmetries that are
broken by scalars with charge $N$ under the respective $U(1)$ acquiring vev's.
However, the presence of a discrete symmetry seems fine-tuned unless there is a
dynamical reason for the scalar field with charge $N$ to acquire an appropriate
vev. 

This might find an explanation within string theory. For example, as has been recently shown in \cite{BerasaluceGonzalez:2011wy}, in type II compactifications discrete symmetries naturally appear as subgroups of anomalous $U(1)$ gauge factors broken by St\"uckelberg type couplings. 
More concretely, D-brane compactifications exhibit multiple $U(1)$'s which generically appear
anomalous whereas the anomalies are cancelled by the Green-Schwarz mechanism \cite{Green:1984ed, Sagnotti:1992qw, Ibanez:1998qp, Bianchi:2000de, Cvetic:2001nr, Antoniadis:2002cs, Anastasopoulos:2003aj, Anastasopoulos:2004ga,  Anastasopoulos:2006cz}. The
$U(1)$'s become massive and are broken to a discrete abelian subgroup via the
presence of a $B \wedge F$ coupling, where $B$ denotes the Ramond Ramond 2-form. 
In the low energy effective theory the massive $U(1)$'s survive as global symmetries
that are preserved by all perturbative quantities while D-instantons can break
them inducing perturbatively absent couplings
\cite{Blumenhagen:2006xt,Ibanez:2006da,Blumenhagen:2009qh}. On the other hand
the discrete symmetry, the remnant of the $U(1)$ after the Green-Schwarz
mechanism, is respected by all perturbative and non-perturbative quantities
\cite{BerasaluceGonzalez:2011wy,Ibanez:2012wg}.

Inspired by the work of \cite{BerasaluceGonzalez:2011wy} we want to investigate the presence of discrete symmetries in
(semi-)realistic D-brane models. In a series of publications
\cite{Cvetic:2009yh,Cvetic:2009ng,Cvetic:2010mm} the authors
analysed so-called D-brane quivers, i.e. local D-brane configurations, with respect
to their phenomenology. They performed a systematic search for local D-brane setups that exhibit (semi-)realistic features using the bottom-up
approach. More concretely, 
they specified the chiral spectrum to be the MSSM or the MSSM with
three right-handed neutrinos and 
imposed the presence of quark and lepton Yukawa couplings on perturbative or
non-perturbative level and at the same time required, among other criteria, the
absence of R-parity violating terms as well as the absence of dimension 5 proton
decay operators. They found of the order of 40 local D-brane configurations
based on four D-brane stacks that are consistent with the global consistency conditions and exhibit a (semi-)realistic phenomenology. Given those
D-brane quivers it is interesting whether the absence of R-parity as well as
dimension 5 proton decay operators is accidental or is originated from a discrete
gauge symmetry.

We will study the constraints arising from string theory for the presence
of a discrete gauge symmetry in D-brane models. As we will see those stringy constraints do
contain the usual four-dimensional discrete anomaly constraints, however, pose
additional constraints related to higher dimensional anomalies upon
decompactification. Established those constraints we analyse the promising local
D-brane configurations found in
\cite{Cvetic:2009yh, Cvetic:2009ng, Cvetic:2010mm} with respect to
discrete gauge symmetries. We find that depending on the hypercharge embedding
matter parity appears quite frequently, forbidding the undesired R-parity
violating terms. Only very few D-brane quivers allow for Proton hexality which
ensures the absence R-parity violating terms as well as the dangerous dimension
5 operators.

This paper is organized as follows. In section \ref{sec discrete field theory}
we review the findings of the systematic search for discrete symmetries in the
MSSM, using four-dimensional discrete anomaly conditions. In section \ref{sec discrete D-brane}
we discuss the constraints on the transformation behaviour of chiral matter
that arise from string consistency conditions. Moreover, we establish the
conditions on the transformation behaviour of the matter fields for the presence
of a discrete gauge symmetry in D-brane compactifications. In section \ref{sec
search} we impose the constraints for the presence of a discrete gauge symmetry,
studied before, for a class of intriguing local D-brane configurations that
exhibit a (semi-) realistic phenomenology. We analyse what type of discrete symmetries
can appear as well as their phenomenological implications. In section \ref{sec
conclusion} we present our conclusions.
The appendix \ref{app bottom up constraints} contains the details of the
systematic bottom-up search for local D-brane configurations that give rise to a (semi-)realistic phenomenology.

\section{Discrete gauge symmetries in the MSSM from a field theory perspective
\label{sec discrete field theory}}

In this section we review the results of the work \cite{Dreiner:2005rd} where
the authors search for all possible family independent (non-R) discrete gauge
symmetries within the MSSM. They find a finite class of discrete gauge
symmetries that satisfy the four-dimensional discrete gauge anomaly constraints, i.e. 
the mixed ${\cal A}_{SU(3) SU(3) \mathbf{Z}_N}$ , ${\cal A}_{SU(2)
SU(2) \mathbf{Z}_N}$ as well as the gravitational anomaly ${\cal A}_{G G \mathbf{Z}_N}$. 
In their search they furthermore require the family independent discrete gauge
symmetries to allow for the Yukawa couplings
\begin{align}
Q_L H_d D_R \qquad \qquad Q_L H_u U_R \qquad \qquad L H_d E_R\,\,.
\end{align}
\begin{table}
\centering
\begin{tabular}{|c|c|c|c|c|c|c|c|c|}
\hline
    & $Q_L$ & $U_R$ & $D_R$ & $L$  & $E_R$ & $N_R$ & $H_u$ & $H_d$\\ 
\hline 
\hline
$A$ & $0$   & $0$   & $-1$  & $-1$ & $0$   & $1$   & $0$   & $1$\\  
$L$ & $0$   & $0$   & $0$   & $-1$ & $1$   & $1$   & $0$   & $0$\\
$R$ & $0$   & $-1$  & $1$   & $0$  & $1$   & $-1$  & $1$   & $-1$\\
\hline
\end{tabular}
\caption{\small The family independent generators of discrete $\mathbf{Z}_N$ gauge symmetries in the MSSM.}
\label{table charges under three Z_N}
\end{table}

As already shown in \cite{Ibanez:1991pr}\footnote{See, also
\cite{Ibanez:1991hv}.} any family independent discrete gauge symmetry 
$\mathbf{Z}_N$ of the MSSM with generator $g_N$ can be expressed in terms of
products of powers of three mutually commuting generators $A_N$, $L_N$ and
$R_N$, i.e.
\begin{align}
g_N = A^{n}_N  \times  L^p_N \times R^m_N\,\,,
\end{align}
where the exponents run over $m, n, p = 0, 1, ... N-1$. The charges of the
chiral MSSM matter fields under these three independent $\mathbf{Z}_N$ are
displayed in table \ref{table charges under three Z_N}. Given this assignment
the matter fields carry discrete charges
\begin{align}
\label{eq discrete charges}
q_{Q_L} = 0   \qquad q_{U_R} =-m   \qquad q_{D_R} = m-n \hspace{20mm}\\  \nn
q_{L}   =-n-p \qquad q_{E_R} = m+p \qquad q_{H_u} = m  \qquad q_{H_d} = -m+n
\end{align}
under a $g_N$ transformation.

The discrete gauge anomaly constraints applying the charge assignment \eqref{eq discrete charges} read 
(see also \cite{Araki:2008ek})
\begin{align}
SU(3) - SU(3) - \mathbf{Z}_N:& \qquad \qquad  3n = 0 \mod N \\
SU(2) - SU(2) - \mathbf{Z}_N:& \qquad \qquad 2n+3p = 0 \mod N\\
G - G - \mathbf{Z}_N:& \qquad \qquad  -13n -3p+3m = 0 \mod N + \eta \frac{N}{2} \,\,,
\label{eq gravitational anomaly}
\end{align}
where the first two lines correspond to the discrete gauge anomalies of $SU(3)$ and $SU(2)$, respectively. The last line describes the gravitational anomaly, where $\eta=0$ for $N$ being odd and $\eta=1$ for $N$ being even. The last term of \eqref{eq gravitational anomaly} takes into account the possibility of heavy Majorana fermion fields. 

Solving these discrete gauge anomaly constraints one finds a finite class
of solutions \cite{Dreiner:2005rd}, ranging from $\mathbf{Z}_2$ up to
$\mathbf{Z}_{18}$ symmetries. In table \ref{table fund} all possible family independent
discrete gauge symmetries of the MSSM are displayed in terms of the three
$\mathbf{Z}_N$ generators, $A_N$, $L_N$ and  $R_N$.

\begin{table}[h]
\center
\hspace{1.00cm}
\scalebox{1.0}{
\begin{tabular}{|c||c|c|c|c|}
\hline
~~~$N$~~~ & ~~~$n$~~~ & ~~~$p$~~~  & ~~~~$m$~~~~&  ~~~~~~~~~Discrete gauge symmetries~~~~~~~~~ \\ 
\hline 
\hline
                  2 & 0 & 0 & 1         & $R_2$\\
\hline
\multirow{2}{*}{3}  & 0 & 0 & 1         & $R_3$\\
\cline{2-5}
                    & 0 & 1 & $(0,1,2)$ & $L_3$, $L_3 R_3$, $L_3 R^2_3$\\
\hline
\multirow{2}{*}{6}  & 0 & 0 & 1         & $R_6$\\
\cline{2-5}
                    & 0 & 2 & $(1,3,5)$ & $L^2_6 R_6$, $L^2_6 R^3_6$, $L^2_6 R^5_6$\\
\hline
\multirow{3}{*}{9}  & 3 & 1 & $(2,5,8)$ & $A^3_9 L_9 R^2_9$, $A^3_9 L_9 R^5_9$, $A^3_9 L_9 R^8_9$\\
\cline{2-5}
                    & 3 & 4 & $(2,5,8)$ & $A^3_9 L^4_9 R^2_9$, $A^3_9 L^4_9 R^5_9$, $A^3_9 L^4_9 R^8_9$\\
\cline{2-5}
                    & 3 & 7 & $(2,5,8)$ & $A^3_9 L^7_9 R^2_9$, $A^3_9 L^7_9 R^5_9$, $A^3_9 L^7_9 R^8_9$\\
\hline
\multirow{3}{*}{18} & 6 & 2 & $(1,7,13)$ & $A^6_{18} L^2_{18} R_{18}$, $A^6_{18} L^2_{18} R^7_{18}$, $A^6_{18} L^2_{18} R^{13}_{18}$\\
\cline{2-5}
                    & 6 & 8 & $(1,7,13)$ & $A^6_{18} L^8_{18} R_{18}$, $A^6_{18} L^8_{18} R^7_{18}$, $A^6_{18} L^8_{18} R^{13}_{18}$\\
\cline{2-5}
                    & 6 & 14 & $(1,7,13)$ & $A^6_{18} L^{14}_{18} R_{18}$, $A^6_{18} L^{14}_{18} R^7_{18}$, $A^6_{18} L^{14}_{18} R^{13}_{18}$\\
\hline
\end{tabular}}
\caption{\small All fundamental discrete gauge symmetries in the MSSM satisfying the 
anomaly cancellation conditions \cite{Dreiner:2005rd}. Here one allows for heavy fermions with fractional charges.}
\label{table fund}
\end{table}

For the MSSM with 3 right-handed neutrinos only a subgroup of the discrete symmetries displayed in table \ref{table fund} can be realized. Requiring the presence of the Dirac neutrino mass $L H_u N_R$ implies the charge
\begin{align}
q_{N_R} = n+p-m
\end{align} 
under the discrete symmetry. Since the neutrinos are not charged under the $SU(3)$ and $SU(2)$ their presence will only lead to changes in the gravitational discrete gauge anomaly, which is then given by
\begin{align}
G - G - \mathbf{Z}_N: \qquad \qquad -10 n =0 \mod N   + \eta \frac{N}{2}  \,\,,
\end{align}
which together with the other two discrete gauge anomaly constraints allows only for solutions with $n=0$. Thus in contrast to the pure MSSM the MSSM  with three additional right-handed neutrinos does not allow any $\mathbf{Z}_9$ and $\mathbf{Z}_{18}$ symmetries. On the other hand all $\mathbf{Z}_2$, $\mathbf{Z}_3$ and $\mathbf{Z}_6$ symmetries displayed in table \ref{table fund} are realized. Beyond those the MSSM with three right-handed neutrinos does not exhibit any further family independent discrete gauge symmetries.

\begin{table}[htb]
\scalebox{1.0}{
\begin{tabular}{|c||c||c|c|c|c||c|c|c|c||c| }
\hline
coupling         & $R_2$      & $ L_3 R_3 $ & $R_3$      & $ L_3$     &$ L_3 R^2_3 $ &$ L^2_6 R^5_6$& $R_6$      &$L^2_6 R^3_6 $&$L^2_6 R_6 $ & \hspace{-1mm}$\mathbf{Z}_9$ \& $\mathbf{Z}_{18}$\hspace{-1.5mm} \\
\hline
\hline
$H_u H_d$        &$\checkmark$&$\checkmark$&$\checkmark$&$\checkmark$&$\checkmark$&$\checkmark$ &$\checkmark$&$\checkmark$ &$\checkmark$&\\
\hline
$L H_u$          &            &$\checkmark$&            &            &            &             &            &             &            &\\
\hline
$L L E_R$        &            &$\checkmark$&            &            &            &             &            &             &            &\\
\hline
$Q_L L D_R$      &            &$\checkmark$&            &            &            &             &            &             &            &\\
\hline
$U_R D_R D_R$    &            &            &            &$\checkmark$&            &             &            &             &            &\\
\hline
$Q_L Q_L Q_L L$  &$\checkmark$&            &$\checkmark$&            &            &             &$\checkmark$&             &            &\\
\hline
\hspace{-0.5mm}$U_R U_R D_R E_R$\hspace{-0.5mm}&$\checkmark$&            &$\checkmark$&            &            &             &$\checkmark$&             &            &\\
\hline
$L H_u L H_u$    &$\checkmark$&$\checkmark$&            &            &            &$\checkmark$ &            &             &            &\\
\hline \hline
$N_R N_R$    &$\checkmark$&$\checkmark$&            &            &            &$\checkmark$ &            &             &            &\\
\hline
\end{tabular}}
\caption{\small Allowed superpotential terms for the respective discrete gauge symmetries \cite{Dreiner:2005rd}.}
\label{table discrete couplings}
\end{table}

For a given discrete gauge symmetry, i.e. for a specific choice of the parameters $m$,
$n$ and $p$, we can determine with eq.~\eqref{eq discrete charges} the discrete
charges of the SM fields and study the appearances of various terms in the
superpotential. Specifically it is interesting whether a discrete symmetry
forbids some of the undesired couplings such as R-parity violating terms or
dangerous dimension 5 proton decay operators. Table \ref{table discrete
couplings} depicts for all possible family independent discrete gauge
symmetries the allowed superpotential terms. 

Let us discuss some of the intriguing discrete symmetries displayed in
table \ref{table discrete couplings}.  
The $\mathbf{Z}_2$ symmetry $R_2$ is the usual matter parity
\cite{Farrar:1978xj} while $L_3 R_3$ is Baryon triality
\cite{Ibanez:1991pr}. Proton hexality, basically the product of matter parity
and Baryon triality, is given by $ L^2_6 R^5_6 $ and forbids all R-parity
violating terms as well as the dangerous dimension 5 proton decay operators while still
allowing for a $\mu$-term $H_u H_d$ and the Weinberg operator $L H_u L H_u$.

The above discussion on the allowed couplings for the respective discrete gauge symmetry applies specifically to the MSSM. Allowing for additional singlets, such as right-handed neutrinos, which do not acquire any vev does not change the analysis. However, the presence of right-handed neutrinos accompanied with a Dirac neutrino mass term raises the issue of the generation of small neutrino masses. A particular intriguing mechanism is the see-saw mechanism that requires large Majorana mass terms for the right-handed neutrinos. In the last line of table \ref{table discrete couplings} we display which of the discrete symmetries permits for a Majorana mass term and thus allows the generation of small neutrino masses via the see-saw mechanism.

Finally, there exist two additional classes of discrete gauge symmetries, namely non-abelian 
discrete gauge symmetries and discrete R-symmetries. As 
recently pointed out the latter may play a special role in GUT theories, realized as a 
$\mathbf{Z}^R_4$ symmetry that forbids all R-parity violating terms as well as dimension 5 proton 
decay operators \cite{Lee:2010gv,Lee:2011dya,Chen:2012jg}. On the other hand non-abelian discrete 
gauge symmetries are often times invoked explaining various observations in flavour physics 
(see e.g. \cite{Altarelli:2010gt}). 
In this work we perform a systematic bottom-up D-brane analysis which ignores any specifics 
of the internal geometry. However non-abelian discrete gauge symmetries as well as discrete 
R-symmetries do rely on the details of the compactification manifold. Thus here we focus 
only on the subset of abelian discrete gauge symmetries.

\section{Discrete symmetries in D-brane compactifications
\label{sec discrete D-brane}}

In this work we want to perform a systematic bottom-up study for discrete gauge 
symmetries within the class of realistic local D-brane configurations based on 
four D-brane stacks found in \cite{Cvetic:2009yh,Cvetic:2009ng,Cvetic:2010mm}
\footnote{The first local bottom-up constructions were discussed in
\cite{Antoniadis:2000ena,Aldazabal:2000sa,Antoniadis:2001np}. For recent analogous 
work on semi-realistic bottom-up searches, see \cite{Ibanez:2008my,Leontaris:2009ci, Kiritsis:2009sf,Anastasopoulos:2010ca,Fucito:2010dk,Cvetic:2010dz,Anastasopoulos:2010hu,Cvetic:2011iq,Cvetic:2012kv,Cvetic:2012kj}.}. 
There the authors studied local D-brane constructions, where the gauge degrees of 
freedom are given by open strings attached to a D-brane stack whereas the chiral 
matter appears at the intersection of two D-brane stacks. As we will review below 
the distribution of the chiral matter is not arbitrary but subject to severe 
constraints arising from string consistency constraints, such as tadpole cancellation. 

In the work \cite{Cvetic:2009yh,Cvetic:2009ng,Cvetic:2010mm} the authors systematically 
analysed all four-stack configurations, four stack quivers, imposed the severe consistency 
constraints as well as required some (semi-) realistic features. More specifically, they 
demanded the chiral spectrum to be one of the MSSM or MSSM plus three right-handed 
neutrinos, required the presence of the quark and lepton Yukawa couplings, the absence 
of R-parity violating superpotential terms as well as the absence of dimension 5 proton 
decay operators. Moreover, in this search they asked for a mechanism that explains the 
small neutrino masses and demanded quark Yukawa textures that are in agreement with the 
CKM matrix. They found of the order of 40 local D-brane configurations allowing for an 
intriguing low energy phenomenology. Here we want to analyse those promising four stack 
quivers with respect to discrete gauge anomalies, i.e. we study whether the absence of 
undesired superpotential terms can be explained by the presence of a discrete gauge symmetry.

Recently discrete gauge symmetries attracted a lot of attention in the construction of 
realistic string theory model building. In the context of heterotic string theory, for 
specific toroidal compactifications one could identify proton hexality, forbidding for 
this specific model the presence of R-parity violating couplings and dimension 5 proton 
decay operators \cite{Forste:2010pf}. Moreover, in \cite{Kappl:2010yu} the authors found for 
a similar construction a $\mathbf{Z}^{R}_4$ symmetry realized that forbids any undesired couplings 
and allows the $\mu$-term only non-perturbatively, thus giving an explanation for the small 
value of around 100 $GeV$. 

Inspired by the work of Banks and Seiberg \cite{Banks:2010zn} discrete gauge symmetries were 
studied also in the context of D-brane compactifications. In \cite{BerasaluceGonzalez:2011wy} 
(see also \cite{Ibanez:2012wg})  abelian discrete gauge symmetries arising from anomalous 
$U(1)$ gauge factors were investigated. This study has been very recently extended to abelian 
and non-abelian discrete gauge symmetries arising from isometries of the compactification 
manifold \cite{BerasaluceGonzalez:2012vb}.

As discussed above we want to perform a systematic study of discrete gauge symmetries within 
a class of local D-brane configurations without making any reference to the details of the 
compactification manifold. Thus in this work we focus on the first class of abelian discrete 
gauge symmetries investigated in \cite{BerasaluceGonzalez:2011wy}. There the authors discuss
the presence of abelian discrete gauge symmetries in D-brane compactifications
which are remnants of anomalous $U(1)$ gauge symmetries which generically
appear in D-brane compactifications. Those anomalous $U(1)$ gauge symmetries
become massive via the Green-Schwarz mechanism and survive as global symmetries 
on the perturbative level. D-instanton effects can break
those global symmetries inducing sometimes desired, but perturbatively
forbidden, couplings, such as Majorana mass terms for the right-handed neutrinos
\cite{Blumenhagen:2006xt,Ibanez:2006da,Cvetic:2007ku,Ibanez:2007rs} or
particular Yukawa couplings in GUT theories \cite{Blumenhagen:2007zk}. As shown
in \cite{BerasaluceGonzalez:2011wy} discrete abelian gauge symmetries in D-brane
compactifications are not broken by non-perturbative effects and thus hold not only at
all levels in perturbation theory, but also at the non-perturbative level.

Here we focus on the concrete case of type IIA constructions with intersecting
D6 branes, but an analogous discussion applies to the T-dual type IIB picture
with D-branes on singularities as well as the type I compactification with magnetized D9 branes.  
In those compactifications D6-branes fill out the four-dimensional space-time
and wrap three-cycles $\pi_x$ in the internal manifold\footnote{For recent reviews on D-brane model building,
see
\cite{Blumenhagen:2005mu,Blumenhagen:2006ci,Marchesano:2007de,Cvetic:2011vz}.
For
original work on globally consistent non-supersymmetric intersecting D-branes, see
\cite{Blumenhagen:2000wh,Aldazabal:2000dg,Aldazabal:2000cn,Blumenhagen:2001te},
and for chiral globally consistent  supersymmetric ones, see
\cite{Cvetic:2001tj,Cvetic:2001nr}. For supersymmetric MSSM realizations, see \cite{Honecker:2003vq,Honecker:2004kb,Gmeiner:2008xq},
and for supersymmetric constructions within type II RCFT's, see \cite{Dijkstra:2004cc,Anastasopoulos:2006da}.
}. A stack of $N$ D6-branes gives
rise to an $U(N)$ gauge theory, that splits into $U(N)= SU(N) \times U(1)$ where
the abelian part is generically anomalous. It becomes massive via the
St\"uckelberg mechanism and does not appear in the low-energy field theory
dynamics. In \cite{Aldazabal:2000sa} the authors give the criteria for the
existence of an unbroken abelian gauge symmetry in the low energy effective theory. For the linear combination
\begin{align}
U(1)= \sum_{x} q_x U(1)_x\,\,,
\end{align}
where the respective $U(1)$ factors originate from the various D-brane stacks,
to remain massless the criterion reads \cite{Aldazabal:2000sa}
\begin{align}
\frac{1}{2}\sum_{x} q_x N_x (\pi_x -\pi'_x) =0 \,\,.
\label{eq massless U(1)}
\end{align}
Here the sum runs over all D-brane stacks $x$ in the given global setup and
$\pi'_x$ denotes the orientifold image cycle of $\pi_x$. 
In order to avoid later confusions in the discussion of discrete symmetries let
us elaborate on constraint \eqref{eq massless U(1)}. 
We introduce a basis of three-cycles $\{\alpha_i\}$ and $\{\beta_i\}$ that are
even and odd under the orientifold action, respectively, with
$i=1,...,h_{21}+1$. The choice of basis is such that $\alpha_i \cdot \beta_j=
\delta_{ij}$ and $\alpha_i \cdot \alpha_j =\beta_i \cdot \beta_j=0$. Then a
three-cycle $\pi_x$ and its orientifold image $\pi'_x$ wrapped by a D-brane stack 
and its image D-brane stack, respectively, can be expanded in terms of this basis 
\begin{align}
\pi_x=\sum_i (m^i_x \alpha_i+ n^i_x \beta_i) \qquad \qquad \pi'_x=\sum_i (m^i_x \alpha_i- n^i_x \beta_i)\,\,,
\label{eq three-cycles}
\end{align}
where $m^i_x$ and $n^i_x$ are integer and are usually referred to as wrapping
numbers. Using eq.~\eqref{eq three-cycles} the constraint \eqref{eq massless U(1)}
takes the form
\begin{align}
\sum_{i} \sum_{x} q_x N_x n^i_x \beta_i = 0\,\,.
\end{align}
Given that the three-cycles $\beta_i$ are orthogonal to each other 
eq.~\eqref{eq massless U(1)} reads
\begin{align}
\sum_{x} q_x N_x n^i_x = 0 \qquad \forall i\,\,.
\end{align}

For a discrete gauge symmetry $\mathbf{Z}_N$ arising from a linear combination
\begin{align}
\mathbf{Z}_N = \sum_{x} k_x U(1)_x
\label{eq def discrete symmetry}
\end{align}
to survive in the low energy effective field theory it has to satisfy
\cite{BerasaluceGonzalez:2011wy}
\begin{align}
\frac{1}{2}\sum_{x} k_x N_x (\pi_x - \pi'_x) = 0 \,\,\,\, \mod N \,\,.
\label{eq discrete condition}
\end{align}
Here we normalize the $k_x$ to be all integer in order to properly identify the
discrete gauge symmetry. Let us clarify the left-hand side of 
eq.~\eqref{eq discrete condition}, which is supposed to indicate that on the left-hand 
side the basis cycles $\beta_i$ appear only in multiples of $N$. More specifically, 
using again the expansion of the three-cycles in terms of the basis $\{\alpha_i\}$ 
and $\{\beta_i\}$ the constraint \eqref{eq discrete condition} reads
\begin{align}
\sum_{x} k_x N_x n^i_x = 0 \mod N \qquad \forall i\,\,,
\label{eq discrete constraint in basis}
\end{align}
In \cite{BerasaluceGonzalez:2011wy,Ibanez:2012wg} the authors give the
constraint for having a discrete symmetry in the form \eqref{eq discrete
constraint in basis} \footnote{In \cite{BerasaluceGonzalez:2011wy} the actual constraint is given by 
\begin{align*}
\sum_x k_x N_x \pi_x \circ \alpha_i = 0 \mod N \qquad \forall i
\end{align*} 
which coincides with \eqref{eq discrete condition} in case $\alpha_i \circ \beta_j= \delta_{ij}$, but differs for the cases where $\alpha_i \circ \beta_j=2\delta_{ij}$ by a factor of $2$. We compensate that by extending the range for the entries $k_x$ to the set $[0,2N-1]$. 
See also, section \ref{sec search}. We thank G. Honecker for pointing out this issue.}. However, for the derivation of stringy bottom-up
constraints the cycle constraints \eqref{eq massless U(1)} and \eqref{eq
discrete condition} will turn out to be more appropriate.

We want to analyse in a bottom-up fashion, consistent with global embedding conditions, i.e. string constraints, analogously to the work
of \cite{Cvetic:2009yh,Cvetic:2009ng,Cvetic:2010mm}\footnote{See
also,\cite{Gmeiner:2005vz,Anastasopoulos:2006da,Anastasopoulos:2010hu}. }
what kind of discrete gauge symmetries do appear in (semi-)realistic D-brane
compactifications. Given a local configuration of D-brane stacks, where the
gauge degrees of freedom are given by open strings localized at a stack of
D-branes while the chiral matter is localized at an intersection of two D-brane
stacks, the chiral matter content cannot be arbitrary. In contrast, it is subject
to severe consistency conditions, namely the tadpole constraint, given by
\begin{align}
\sum_x N_x \left( \pi_x + \pi'_x\right) = 4 \pi_{O6}\,\,,
\label{eq tadpole}
\end{align}
as well as the constraint \eqref{eq massless U(1)} required for
the presence of a massless $U(1)$ in the low energy effective theory. Here
$\pi_{O6}$ in eq.~\eqref{eq tadpole} denotes the homology class of the orientifold
plane.

\begin{table}
\centering
\begin{tabular}{|c|c|}
\hline
Representation                   & Multiplicity \\
\hline 
\hline 
$\Ysymm_a$                       & $\#(\Ysymm_a)            =  {1\over 2}\left(\pi_a\circ \pi'_a-\pi_a \circ  \pi_{{\rm O}6} \right)$\\
$\Yasymm_a$                      & $\#(\Yasymm_a)           =  {1\over 2}\left(\pi_a\circ \pi'_a+\pi_a \circ  \pi_{{\rm O}6} \right)$\\
$( \fund_a,{\overline \fund}_b)$ & $\#(\fund_a, \ov \fund_b)= \pi_a\circ \pi_{b}$\\
$(\fund_a, \fund_b)$             & $\#(\fund_a,\fund_b)     = \pi_a\circ \pi'_{b}$\\
\hline
\end{tabular}
\vspace{2mm} 
\caption{\small Chiral spectrum of intersection D-branes.}
\label{table chiral spectrum}
\end{table}

 The equations \eqref{eq tadpole} and \eqref{eq massless U(1)} are
conditions on the three-cycles the D6-branes wrap, and imply the
transformation behaviour of the four-dimensional chiral matter under the D-brane
gauge symmetries. More specifically, the chiral matter fields cannot be
distributed arbitrarily at the intersections of stacks of D-branes, but they have to obey the above conditions.
Those constraints on the transformation
behaviour of the  matter fields under the D-brane gauge symmetries can be
derived by multiplying the equations \eqref{eq tadpole} and \eqref{eq massless
U(1)} with the three-cycles wrapped by the D6-branes and using table \ref{table
chiral spectrum}.

From the tadpole constraint one obtains  \cite{Anastasopoulos:2006da,Cvetic:2009yh,Anastasopoulos:2010hu}
\begin{align}
\sum_{x \neq a} N_x \Big( \#(\fund_a, \ov \fund_x) +  \#(\fund_a, \fund_x) \Big) + (N_a - 4)\#( \, \Yasymm_a) + (N_a + 4) \#(\Ysymm_a) = 0 \,\,,
\label{eq constraint1}
\end{align}
which is a constraint for each D-brane stack $a$ of the D-brane setup. Due to
the absence of antisymmetric representations for abelian gauge symmetries for a
$U(1)$ stack, for a single D-brane stack, the constraint takes the form
\begin{align}
\sum_{x \neq a} N_x \Big( \#(\fund_a, \ov \fund_x) +  \#(\fund_a, \fund_x) \Big) + 5 \# (\Ysymm_a)=0  \qquad \text{mod} \,3\,\,.
\label{eq constraint2}
\end{align}
Note that eq.~\eqref{eq constraint1} is exactly the anomaly cancellation condition
for non-abelian gauge symmetries\footnote{This statement holds true for all $SU(N)$ with $N>2$. For  $SU(2)$ field theory does not distinguish between ${\mathbf 2}  = \ov {\mathbf 2} $, however string theory knows about the $U(2)$ origin of $SU(2)$ and thus differentiates between ${\mathbf 2} = \ov{\mathbf 2} $.}. However, condition \eqref{eq constraint2} has
no four-dimensional field theory analogue. It should be stressed that the
constraints \eqref{eq constraint1} and \eqref{eq constraint2} are only necessary
constraints. A given tadpole free chiral spectrum arising from a local D-brane
configuration does satisfy eq.~\eqref{eq constraint1} and \eqref{eq constraint2}.
However, a spectrum satisfying eq.~\eqref{eq constraint1} and \eqref{eq constraint2}
is not necessarily tadpole free.

The constraints on the transformation behaviour of matter field for having an
abelian gauge symmetry in the low energy effective action arising from eq.~\eqref{eq
massless U(1)} takes the form\cite{Cvetic:2009yh,Anastasopoulos:2010hu}
\begin{align}
\frac{1}{2}\sum_{x \neq a} q_x\, N_x \#(\fund_a,{\ov \fund_x}) 
& - \frac{1}{2}  \sum_{x \neq a} q_x\, N_x \#(\fund_a,\fund_x)
\label{eq massless constraint abelian}\\
& \hspace{-10mm} = \frac{q_aN_a}{2(4-N_a)}\,\left(\sum_{x \neq a} N_x \Big( \#(\fund_a, \ov \fund_x) +  \#(\fund_a, \fund_x)  \Big) + 8 \#(\Ysymm_a)\right)  \,\,,\nn
\end{align}
where we multiplied equation \eqref{eq massless U(1)} with the homology class of the three-cycles 
wrapped by the D-brane stack $a$. In the derivation of eq.~\eqref{eq massless
constraint abelian} we used \eqref{eq constraint1} to eliminate the
anti-symmetrics. That allows us to display the constraints on the transformation
behaviour of the matter fields independently of whether the considered D-brane
stack consists of a single or multiple D6-branes. As for the tadpole constraint
one has one constraint for each D-brane stack $a$. Moreover, the constraints
\eqref{eq massless constraint abelian} do imply the cancellation of abelian cubic
anomalies as well as mixed anomalies in four dimensions. However, the
constraints imply additional conditions on the transformation behaviour of 
chiral matter beyond four-dimensional abelian gauge anomaly cancellation. The
additional constraints are related to the cancellation of higher dimensional
anomalies upon decompactification \cite{Ibanez:1998qp, Antoniadis:2002cs, Anastasopoulos:2003aj}.

Again we would like to stress that condition \eqref{eq massless constraint
abelian} is only a necessary constraint, but not sufficient. This means that not
any chiral spectrum arising from a local D-brane configuration satisfying
\eqref{eq massless constraint abelian} for a linear combination of the abelian
$U(1)$ factors does exhibit this abelian gauge symmetry in the low energy
effective action.  

Let us assume that the MSSM is realized only on a subset of D6-brane stacks, and
furthermore that the remaining D6-branes, so called hidden sector D-branes whose presence may be required for
global consistency, do not intersect the MSSM D6-branes chirally. For such a
scenario the sum over $x$ in equations \eqref{eq constraint1}, \eqref{eq
constraint2} and \eqref{eq massless constraint abelian} does only contain the
visible MSSM D-brane stacks and no knowledge of the hidden sector 
is necessary.

In the work \cite{Cvetic:2009yh,Cvetic:2009ng,Cvetic:2010mm} the
authors investigated various local D-brane configurations by specifying for each
local setup the origin of the chiral MSSM matter, i.e. they specify the
intersection at which the chiral matter fields do appear, and systematically
analyse whether such local configurations do satisfy the severe constraints laid
out above. They furthermore required a set of phenomenological bottom-up
constraints to ensure compatibility with experimental observations. The latter
contain among others the absence of R-parity violating couplings on the
perturbative level. We summarize them in appendix \ref{app bottom up
constraints}.

Here we want to investigate whether those promising local D-brane configurations
found in \cite{Cvetic:2009yh,Cvetic:2009ng,Cvetic:2010mm}, that are consistent with the global consistency conditions, do
exhibit discrete gauge symmetries and analyse their phenomenological
implications for the low energy effective field theory. 

The condition to have a discrete symmetry in a D-brane compactification
\eqref{eq discrete condition} is a constraint on the three-cycles the D6-branes
wrap. Just as for the tadpole constraint \eqref{eq tadpole} and the masslessness
condition \eqref{eq massless U(1)} we will translate this cycle condition into a
constraint on the transformation behaviour of the chiral matter by multiplying
eq.~\eqref{eq discrete condition} with the homology class of the three-cycles wrapped by
the MSSM branes and apply table \ref{table chiral spectrum}. One obtains 
\begin{align}
\frac{1}{2} \sum_{x \neq a} k_x\, N_x \Big( \#(\fund_a,{\ov \fund_x})- \#(\fund_a,\fund_x)\Big) 
\label{eq massless constraint discrete}
& - \frac{k_a N_a}{2}\Big(\#(\Ysymm_a) + \#(\Yasymm_a)\Big) = 0 \mod N\,, 
\end{align}
which represents a separate constraint for each D-brane stack $a$. Note that due to
the non-integer prefactor $\frac{1}{2}$ in equation \eqref{eq massless
constraint discrete} the $k_x$ do lie in the interval $(0, 2N-1)$. Furthermore, 
for $U(1)$ D-brane stacks there are no massless antisymmetrics. In an
analogous fashion as for the massless $U(1)$'s condition (see eq. \eqref{eq massless constraint abelian}) we use the tadpole
condition \eqref{eq constraint1} to eliminate the antisymmetrics that results
into
\begin{align}
\frac{1}{2}\sum_{x \neq a} k_x\, N_x \#(\fund_a,{\ov \fund_x}) &- \frac{1}{2}  \sum_{x \neq a} k_x\, N_x \#(\fund_a,\fund_x)
\label{eq discrete constraint abelian}\\
& \hspace{-35mm}- \frac{k_aN_a}{2(4-N_a)}\,\left(\sum_{x \neq a} N_x \Big(\#(\fund_a, \ov \fund_x) +  \#(\fund_a, \fund_x)  \Big) + 8 \#(\Ysymm_a)\right)  =0\,\, \mod N \nn\,\,.
\end{align}
One has to be slightly careful in using the tadpole constraint \eqref{eq
tadpole} to replace the antisymmetrics due to the fact that generically the
prefactor is non-integer. This is not an issue for the presence of an abelian
gauge symmetry, however can be very crucial for discrete gauge symmetries since
the left hand side is not $0$ but rather $0 \mod N $. One can compensate that by
enlarging the interval for the $k_x$ or by requiring an additional constraint
arising from multiplying the homology class of the orientifold plane with the discrete
symmetry constraint \eqref{eq discrete condition} \footnote{Note that for the
abelian gauge symmetry such an additional constraint is not necessary, since one can use
the tadpole constraint to replace the homology class of the orientifold plane by
all the three-cycles wrapped by the D-brane stacks.}. This additional constraint
reads
\begin{align}
\sum_a k_a N_a \Big( \#(\Yasymm_a) - \#(\Ysymm_a)\Big) = 0 \mod N\,\,,
\end{align}
which after replacing the antisymmetrics in order not to have to distinguish
between non-abelian and abelian D-brane stacks takes the form
\begin{align}
\sum_a \frac{k_a N_a}{4-N_a} \left( \sum_{x \neq a} N_x \Big( \#(\fund_a, \ov \fund_x) + \#(\fund_a, \fund_x)  \Big)  + 2 N_a\#(\Ysymm_a) \right) = 0 \mod N\,\,.
\label{eq discrete additional}
\end{align}
 Let us mention that the constraints \eqref{eq discrete constraint abelian} and
\eqref{eq discrete additional} do imply the vanishing of the various discrete
gauge anomalies, such as $ SU(N) - SU(N) - \mathbf{Z}_N$ or $G -G - \mathbf{Z}_N $. 
However, analogously to the abelian gauge symmetry these 
string theory constraints are more severe than just four-dimensional discrete gauge
anomaly cancellation.

In the subsequent chapter we will investigate the quivers, local D-brane
configurations, that were found in
\cite{Cvetic:2009yh,Cvetic:2009ng,Cvetic:2010mm}  with respect to
discrete symmetries. We will analyse to what extend discrete gauge symmetries do
arise and their implications for the low energy action. We will compare those
discrete symmetries with the ones found in a pure field theory context
\cite{Ibanez:1991hv,Ibanez:1991pr,Dreiner:2005rd}.

\section{Systematic bottom-up search\label{sec search}}
In the work \cite{Cvetic:2009yh,Cvetic:2009ng,Cvetic:2010mm} the
authors found various local D-brane configuration, containing up to four D-brane stacks, giving rise to
the MSSM spectrum and extensions of it that satisfy the severe top-down
constraints arising from string theory, see equations \eqref{eq constraint1},
\eqref{eq constraint2} and \eqref{eq massless constraint abelian} as well as
some minimal set of phenomenological requirements, so called bottom-up
constraints. Those contain constraints on R-parity violating couplings, on
dangerous dimension 5 proton decay operators as well as on Yukawa textures. In
appendix \ref{app bottom up constraints} we summarize those phenomenological
bottom-up constraints. 

Here we will study those local D-brane configurations, that are consistent with the global consistency conditions, with respect to discrete
symmetries. We will analyse what quivers do satisfy the constraints to exhibit
discrete symmetries and investigate their implications on the superpotential
couplings.

Let us lay out the details of the search. For a chosen $N$ we check whether a
given linear combination of $U(1)$'s in terms of the vector $(k_a, k_b, k_c,
k_d)$, with the $k_x$'s being integers, does satisfy the constraints \eqref{eq discrete constraint abelian} and
\eqref{eq discrete additional}. Due to the prefactor $\frac{1}{2}$ in eq.~\eqref{eq
discrete condition} we let the $k_x$ run from $0$ to $2N-1$. 

Via a hypercharge shift we can find to any given solution $(k_a, k_b, k_c, k_d)$
an additional equivalent solutions by adding the hypercharge. Thus  $(k_a +m
y_a, k_b+ m y_b, k_c+ m y_c, k_d+m  y_d)$ is also a solution to the constraints
\eqref{eq discrete constraint abelian} and \eqref{eq discrete additional} where
$m$ is an integer and the $y_x$ denote the integer hypercharge embedding
coefficients. In order to avoid overcounting we fix the discrete charge of $Q_L$
for one family to be $0$ by choosing $k_a=k_b$\footnote{In our displayed local
D-brane configurations at least one of the left-handed quarks transforms as
$(\fund_a, \overline \fund_b)$ under the D-brane gauge symmetry  $U(3)_a \times
U(2)_b$.}. Thus we run only over three free integer parameter, namely $k_a$,
$k_c$ and $k_d$.

Additionally, we demand that the discrete symmetries allow for the quark and
lepton Yukawa couplings in the superpotential, whose presence is crucial for the
generation of low energy fermion masses. It turns out that this requirement is
very stringent and rules out various discrete symmetries which otherwise satisfy
the discrete top-down constraints \eqref{eq discrete constraint abelian} and
\eqref{eq discrete additional}.

Finally, we often find solutions for discrete gauge symmetries of higher
degree due to the $\frac{1}{2}$ in \eqref{eq discrete constraint abelian} and
\eqref{eq discrete additional}, such as $\mathbf{Z}_{12}$, which eventually after determining the matter
field charges turn out to be of lower degree from a pure MSSM point of view, since all matter charges have a
common divisor. We take those things into account when identifying the discrete
symmetries but nevertheless display the linear combinations describing the
discrete gauge symmetries in the D-brane language. Therefore, it frequently happens
that $\mathbf{Z}_6$ symmetries contain coefficients that are higher than $12$.

In the following we investigate the various promising four-stack quivers, which
give rise to the MSSM spectrum (see section \ref{MSSM}) and the MSSM spectrum with
three right-handed neutrinos (see section \ref{sec MSSM + 3N}). Those
promising quivers, that are consistent with the global consistency conditions, were found in a systematic bottom-up search performed in
\cite{Cvetic:2009yh,Cvetic:2009ng,Cvetic:2010mm}. In sections \ref{MSSM}  and \ref{sec MSSM + 3N} we give the details of our findings, specifically we display for each D-brane configuration the possible discrete symmetries and their corresponding vectors $(k_a, k_b, k_c, k_d)$. In section \ref{sec summary} we present a broad summary of our results.

\subsection{MSSM realizations
\label{MSSM}}
Here we investigate all four stack realizations that give rise to the exact MSSM
spectrum, satisfy the severe top-down constraints discussed above and pass the
phenomenological constraints displayed in appendix \ref{app bottom up
constraints}. Those quivers were found in the systematic bottom-up search
performed in \cite{Cvetic:2010mm}. There the authors found D-brane
configurations for two different hypercharge embeddings, namely 
\begin{itemize}
\item[$\bullet$] $U(1)_Y= -\frac{1}{3}\, U(1)_a - \frac{1}{2} \,U(1)_b + U(1)_d$
\item[$\bullet$] $U(1)_Y= -\frac{1}{3}\, U(1)_a - \frac{1}{2} \,U(1)_b  $
\end{itemize} 
which we will discuss subsequently.

\subsubsection{Hypercharge $U(1)_Y= -\frac{1}{3} U(1)_a - \frac{1}{2} U(1)_b + U(1)_d $}
For this hypercharge embedding there are three different D-brane quivers that give
rise to realistic phenomenology. They are displayed in table \ref{four stack
two}. 
\begin{table}[h]
\centering
\scalebox{.90}{
\begin{tabular}{|c|c|c|c|c|c|c|c|c|c|c|}\hline
\multirow{2}{*}{$\#$}&\multicolumn{1}{|c}{$Q_L$} & \multicolumn{1}{|c}{$D_R$} & \multicolumn{2}{|c}{$U_R$} & \multicolumn{1}{|c}{$L$} & \multicolumn{2}{|c}{$E_R$} & \multicolumn{1}{|c}{$H_u$} & \multicolumn{2}{|c|}{$H_d$}\\
\cline{2-11}
&$(\fund_a,\ov\fund_{b})$ 
&$(\ov\fund_{a},\ov \fund_{c})$
&$(\ov \fund_{a},\ov \fund_{d})$ & $\Yasymm_a$
&$(\fund_b,\ov\fund_{c})$
&$(\fund_c,\fund_d)$ & $\ov{\Yasymm}_b$ $\phantom{I^{I^{I^I}}}$\hspace{-8mm} 
&$(\fund_b,\fund_d)$
&$(\fund_b,\fund_c)$ & $(\ov \fund_{b},\ov\fund_{d})$
\\
\hline
\hline
1&3&3&3&0&3&1&2&1&1&0\\
\hline
2&3&3&3&0&3&0&3&1&0&1\\
\hline
3&3&3&0&3&3&0&3&1&0&1\\
\hline
\end{tabular}}
\caption{\small MSSM spectrum for setups with $U(1)_Y= -\frac{1}{3}\, U(1)_a - \frac{1}{2} \,U(1)_b + U(1)_d $.
\label{four stack two}
}
\end{table}
The first two solutions of table \ref{four stack two} exhibit a discrete
$\mathbf{Z}_3$ symmetry  which can be identified with $L_3 R_3$, i.e. Baryon
triality. The linear combination is given by
\begin{align}
L_3 R_3 = U(1)_a + U(1)_b +  3 U(1)_c + U(1)_d
\end{align}
and satisfies the constraints \eqref{eq discrete constraint abelian} and
\eqref{eq discrete additional}. Thus both models may exhibit a Baryon triality. 

The third solution of table \ref{four stack two} may even have an additional
massless $U(1)$ given by 
\begin{align}
U^{add}(1) = U(1)_d\,\,.
\end{align} 
However, it should be absent since otherwise it would spoil the presence of
desired Yukawa couplings. Even any discrete subgroup of $U^{add}(1)$ is
forbidden, since it does not allow any of the superpotential terms $Q_L H_u
U_R$, $Q_L H_d D_R$ nor $L H_d E_R$. Apart from this additional undesired
$U^{add}(1)$ and its potential undesired discrete subgroups this local D-brane
setup does not exhibit any further discrete symmetries.  

\subsubsection{Hypercharge $U(1)_Y= -\frac{1}{3} U(1)_a - \frac{1}{2} U(1)_b$}
\begin{table}[h]
\centering
\scalebox{1.0}{
\begin{tabular}{|c|c|c|c|c|c|c|c|}\hline
\multirow{2}{*}{$\#$}&\multicolumn{1}{|c}{$Q_L$} & \multicolumn{1}{|c}{$D_R$} & \multicolumn{1}{|c}{$U_R$} & \multicolumn{1}{|c}{$L$} & \multicolumn{1}{|c}{$E_R$} & \multicolumn{1}{|c}{$H_u$} & \multicolumn{1}{|c|}{$H_d$}\\
\cline{2-8}
&$(\fund_a,\ov\fund_{b})$ 
&$(\ov\fund_{a},\ov\fund_{d})$
&$\Yasymm_a$
&$(\fund_b,\ov \fund_d)$
&$\ov{\Yasymm}_b$ $\phantom{I^{I^{I^I}}}$\hspace{-8mm} 
&$(\ov\fund_b,\ov \fund_c)$
&$(\fund_b,\fund_c)$
\\
\hline
\hline
1&3&3&3&3&3&1&1\\
\hline
\end{tabular}}
\caption{\small MSSM spectrum for setups with $U(1)_Y = -\frac{1}{3} U(1)_a -\frac{1}{2} U(1)_b$.
\label{four stack three}
}
\end{table}
The solution \# 1 displayed in table \ref{four stack three} may exhibit an additional
$U^{add}(1)$ that is given by
\begin{align}
U^{add}(1) = U(1)_c
\end{align}
satisfying the necessary constraints \eqref{eq massless constraint abelian}. However, such an abelian gauge symmetry and any discrete subgroup
of it should be absent, since otherwise the desired Yukawa couplings $Q_L H_u
U_R$, $Q_L H_d D_R$ and $L H_d E_R$ would be forbidden.

Apart from this additional $U^{add}(1)$ we find no vector $(k_a, k_b, k_c, k_d)$
 that satisfies the discrete anomaly constraints \eqref{eq discrete constraint
abelian} and \eqref{eq discrete additional}. Thus in such a configuration one
cannot find any discrete gauge symmetry, which may help to explain the absence
of various undesired superpotential terms and the absence of R-parity violating
terms or dimension 5 proton decay operators is rather accidental.

\subsection{MSSM + three right-handed neutrino realizations\label{sec MSSM + 3N}}
Here we analyse all four stack realizations exhibiting the MSSM spectrum plus
three right-handed neutrinos, satisfying the severe top-down constraints and
allowing for an acceptable phenomenology. Those quivers were found in a
systematic search performed in \cite{Cvetic:2009yh,Cvetic:2009ng}, where the
authors found only four solutions for the hypercharge embeddings listed below
\begin{itemize}
\item[$\bullet$] $U(1)_Y=  \frac{1}{6} U(1)_a +\frac{1}{2} U(1)_c -\frac{3}{2}U(1)_d$
\item[$\bullet$] $U(1)_Y= -\frac{1}{3} U(1)_a -\frac{1}{2} U(1)_b$
\item[$\bullet$] $U(1)_Y= -\frac{1}{3} U(1)_a -\frac{1}{2} U(1)_b + U(1)_d$
\item[$\bullet$] $U(1)_Y=  \frac{1}{6} U(1)_a +\frac{1}{2} U(1)_c -\frac{1}{2}U(1)_d$\,\,.
\end{itemize}
They will be analysed with respect to discrete gauge symmetries in the following.

\subsubsection{Hypercharge $U(1)_Y= \frac{1}{6} U(1)_a +\frac{1}{2} U(1)_c - \frac{3}{2}U(1)_d$}

\begin{table}[h]
\centering
\scalebox{.95}{
\begin{tabular}{|c|c|c|c|c|c|c|c|c|}\hline
\multirow{2}{*}{$\#$}&\multicolumn{1}{|c}{$Q_L$} & \multicolumn{1}{|c|}{$D_R$} & $U_R$ & \multicolumn{1}{|c}{$L$} & \multicolumn{1}{|c}{$E_R$} & \multicolumn{1}{|c|}{$N_R$} & \multicolumn{1}{|c|}{$H_u$}  & \multicolumn{1}{|c|}{$H_d$} \\
\cline{2-9}
&$(\fund_a,\ov \fund_{b})$  & ${\Yasymm}_a$ & $(\ov\fund_{a},\ov \fund_{c})$ & $(\fund_b,\ov \fund_{c})$ 
& $(\ov \fund_{c},\ov \fund_{d})$   & $\ov{\Yasymm}_b$ $\phantom{I^{I^{I^I}}}$\hspace{-8mm} 
& $(\fund_b,\fund_c)$ & $(\ov \fund_{b},\ov \fund_{c}) $\\
\hline
\hline
1&3&3&3&3&3&3&1&1\\
\hline
\end{tabular}}
\caption{\small MSSM + 3 $N_R$ spectrum for setups with $U(1)_Y=\frac{1}{6}U(1)_a+\frac{1}{2}U(1)_c-\frac{3}{2} U(1)_d$.
\label{table spectrum four stack model 1/6a+1/2c-3/2d}}
\end{table}\vspace{5pt}
In the model \# 1 of table \ref{table spectrum four stack model 1/6a+1/2c-3/2d} the matter fields transform in such a way that there may be an
additional $U^{add}(1)$ given by the linear combination
\begin{align}
U^{add}(1)= U(1)_a+U(1)_b + U(1)_c -3 U(1)_d\,\,,
\end{align} 
which allows for all the desired Yukawa couplings, and together with the hypercharge gives the B-L symmetry:  $U(1)_{B-L}=   
2 U(1)_Y - \frac{1}{2} U^{add}(1)$. Clearly any discrete subgroup of the gauge symmetry $U^{add}(1)$ will satisfy the
constraints \eqref{eq discrete constraint abelian} and \eqref{eq discrete
additional}. The $\mathbf{Z}_2$, $\mathbf{Z}_3$ and $\mathbf{Z}_6$ discrete
subgroups of $U^{add}(1)$ correspond to  $R_2$, $R_3$ and $R_6$, respectively.
Beyond the discrete gauge subgroups of $U^{add}(1)$ the local setup does not
exhibit any additional discrete gauge symmetries.

\subsubsection{Hypercharge $U(1)_Y= -\frac{1}{3} U(1)_a -\frac{1}{2} U(1)_b$}

\begin{table}[h]
\begin{center}
\scalebox{.70}{
\begin{tabular}{|c|c|c|c|c|c|c|c|c|c|c|c|c|c|c|c|}
\cline{1-15}
\multirow{2}{*}{$\#$} & \multicolumn{1}{|c}{$Q_L$} & \multicolumn{1}{|c|}{$D_R$} & \multicolumn{1}{|c|}{$U_R$} & \multicolumn{1}{|c}{$L$} & \multicolumn{1}{|c}{$E_R$} & \multicolumn{7}{|c|}{$N_R$} & \multicolumn{1}{|c|}{$H_u$} & \multicolumn{1}{|c|}{$H_d$} \\
\cline{2-15}
& $(\fund_a,\ov \fund_{b})$ 
& $(\ov \fund_{a},\ov \fund_{d})$ 
& ${\Yasymm}_a$ 
& $(\fund_b,\ov \fund_{d})$ 
& $\ov{\Yasymm}_b$ $\phantom{I^{I^{I^I}}}$\hspace{-8mm} 
& $(\fund_c,\fund_d)$ & $(\ov\fund_{c},\fund_d)$ & $(\ov\fund_{c},\ov \fund_{d})$ & ${\Ysymm}_c$ &$\ov{\Ysymm}_c$ & ${\Ysymm}_d$ & $\ov{\Ysymm}_d$ 
& $(\ov\fund_{b},\ov \fund_{c})$ 
& $(\fund_b,\fund_c)$ 
\\
\hline
\hline
1&3&3&3&3&3&0&0&0&0&0&0&3&1&1\\ \hline
2&3&3&3&3&3&0&0&0&0&0&3&0&1&1\\ \hline
3&3&3&3&3&3&0&0&3&0&0&0&0&1&1\\ \hline
4&3&3&3&3&3&0&3&0&0&0&0&0&1&1\\ \hline
5&3&3&3&3&3&1&1&0&0&0&0&1&1&1\\ \hline
6&3&3&3&3&3&3&0&0&0&0&0&0&1&1\\ \hline
7&3&3&3&3&3&0&0&1&0&1&0&1&1&1\\ \hline
8&3&3&3&3&3&0&1&0&0&1&1&0&1&1\\ \hline
9&3&3&3&3&3&0&0&0&0&3&0&0&1&1\\ \hline
10&3&3&3&3&3&0&1&1&1&0&0&0&1&1\\ \hline
11&3&3&3&3&3&1&0&0&1&0&1&0&1&1\\ \hline
12&3&3&3&3&3&0&0&0&3&0&0&0&1&1\\ \hline
\end{tabular}
}
\caption{\small MSSM + 3 $N_R$ spectrum for setups with $U(1)_Y= -\frac{1}{3} U(1)_a -\frac{1}{2} U(1)_b$.
\label{table four stack model  1/3a - 1/2 c}}
\end{center}
\end{table}
The solutions \# 1, \# 3 and \# 5 of table \ref{table four stack model  1/3a - 1/2 c} may
exhibit an additional $U(1)$, satisfying the constraints \eqref{eq massless constraint abelian}. However, it should be
noted that those $U(1)$'s cannot be realized in a realistic compactification,
since their presence would forbid some of the desired Yukawa couplings $Q_L H_u
U_R$, $Q_L H_d D_R$ or $L H_d E_R$. Not even a discrete subgroup of those
abelian gauge symmetries is allowed since for any discrete subgroup the absence
of the desired Yukawa couplings holds true.

The solution \# 2 may even exhibit two independent abelian gauge symmetries, namely
\begin{align}
U^{add}_1(1) = U(1)_c \qquad \qquad \text{and} \qquad \qquad  U^{add}_2(1) = U(1)_b -2 U(1)_d
\end{align}
Only the linear combination $U^f(1)=-3U(1)_Y + U^{add}_1(1) -\frac{1}{2}
U^{add}_2(1)$ should be indeed realized, since otherwise various desired Yukawa
couplings would be not allowed. This implies the presence of a B-L symmetry that is given by $U(1)_{B-L}=   
2 U(1)_Y + \frac{1}{2} U^f(1)$. Again the constraints \eqref{eq massless
constraint abelian} are only necessary constraints and not sufficient, but
clearly any subgroup of $U^f(1)$ does satisfy the discrete anomaly constraints
\eqref{eq discrete constraint abelian} and \eqref{eq discrete additional}. Thus
even though $U^f(1)$ may not be realized in a concrete compactification it may
well be that a discrete subgroup survives. Among those subgroups rank the
$\mathbf{Z}_2$, $\mathbf{Z}_3$ and $\mathbf{Z}_6$ discrete symmetries $R_2$
$R_3$ and $R_6$.  

In addition to the above mentioned observations we find for all solutions apart
for solutions \# 4, \# 5, \# 8 and \# 10 the discrete symmetry $R_2$ realized, where the
matter field charges are given by
\begin{align}
R_2 = U(1)_a + U(1)_b + U(1)_c + U(1)_d\,\,.
\end{align}
As can be seen from table \ref{table fund} this matter parity $R_2$
forbids the presence of R-parity violating couplings. Beyond matter parity $R_2$
none of the twelve setups exhibits any additional discrete gauge symmetries,
apart from solution \# 2 which allows for discrete subgroups of $U^{f}(1)$.

\subsubsection{Hypercharge $U(1)_Y= -\frac{1}{3}U(1)_a -\frac{1}{2} U(1)_b + U(1)_d$}

\begin{table}[h]
\centering
\scalebox{.80}{
\begin{tabular}{|c|c|c|c|c|c|c|c|c|c|c|c|c|} \hline
\multirow{2}{*}{$\#$} & \multicolumn{1}{|c}{$Q_L$} & \multicolumn{1}{|c|}{$D_R$} & \multicolumn{2}{|c|}{$U_R$} & \multicolumn{1}{|c}{$L$} & \multicolumn{2}{|c}{$E_R$} & \multicolumn{2}{|c|}{$N_R$} & \multicolumn{1}{|c|}{$H_u$} & \multicolumn{2}{|c|}{$H_d$} \\
\cline{2-13}
& $(\fund_a,\ov \fund_{b})$ 
& $(\ov \fund_{a},\ov \fund_{c})$ 
& $(\ov\fund_{a},\ov\fund_{d})$ & ${\Yasymm}_a$ 
& $(\fund_b,\ov\fund_{c})$ 
& $(\fund_c,\fund_d)$ & $\ov{\Yasymm}_b$ $\phantom{I^{I^{I^I}}}$\hspace{-8mm} 
& ${\Ysymm}_c$ & $\ov{\Ysymm}_c$ 
& $(\fund_b,\fund_d)$ 
& $(\fund_b,\fund_c)$ & $(\ov\fund_{b},\ov\fund_{d})$ 
\\
\hline
\hline
1&3&3&3&0&3&1&2&0&3&1&1&0\\ \hline
2&3&3&3&0&3&1&2&3&0&1&1&0\\ \hline
3&3&3&0&3&3&0&3&0&3&1&0&1\\ \hline
4&3&3&0&3&3&0&3&3&0&1&0&1\\ \hline
\end{tabular}}
\caption{\small MSSM + 3 $N_R$ spectrum for setups with $U(1)_Y= -\frac{1}{3} U(1)_a -\frac{1}{2} U(1)_b + U(1)_d $.
\label{table four stack model  1/3a - 1/2 b+1 d}
}
\end{table}
The solution \# 1 of table \ref{table four stack model  1/3a - 1/2 b+1 d} satisfies
all constraints for matter parity, Baryon triality and hence also for Proton hexality.
Matter parity $R_2$
and Baryon triality $L_3 R_3$ are given by
\begin{align}
R_2     & = U(1)_a + U(1)_b + U(1)_c + 5 U(1)_d \\
L_3 R_3 & = U(1)_a + U(1)_b + 3 U(1)_c + U(1)_d \,\,.
\end{align}
Proton hexality takes the form
\begin{align}
L^2_6 R^5_6 = U(1)_a + U(1)_b + 9 U(1)_c + 13 U(1)_d\,\,
\end{align}
and does prevent the presence of R-parity violating couplings as well as the
presence of dangerous dimension 5 proton decay operators, and at the same time
allows for a $\mu$-term as well as a Weinberg operator.

The second solution of table \ref{table four stack model  1/3a - 1/2 b+1 d}
exhibits a massless $U(1)$ of the form 
\begin{align}
U^{add}(1)=U(1)_a +U(1)_b +U(1)_c -3U(1)_d  
\end{align}
which does not forbid any desired Yukawa couplings whereas the B-L symmetry takes the form
$U(1)_{B-L} = 2 U(1)_Y + \frac{1}{2} U^{add}(1)$. As before any discrete
subgroup satisfies the constraints for the discrete symmetry \eqref{eq discrete
constraint abelian} and \eqref{eq discrete additional}. For instance the
$\mathbf{Z}_2$ subgroup of $U^{add}$ can be interpreted as matter parity.
Moreover, one finds all four different discrete $\mathbf{Z}_3$  symmetries found
in the MSSM using the pure field theoretical ansatz. They are given by the
following linear combinations
\begin{align}
L_3 R^2_3 & = 2U(1)_c + 4U(1)_d \\  
L_3       & = U(1)_a + U(1)_b + 5U(1)_c +5U(1)_d\\
R_3       & = U(1)_a + U(1)_b +U(1)_c + 3U(1)_d\\
L_3 R_3   & = U(1)_a + U(1)_b +3 U(1)_c + U(1)_d\,\,,
\end{align}
where $R_3$ originates from $U^{add}(1)$. Only Baryon triality $L_3 R_3$ allows for the presence of a Weinberg operator.
Thus in presence of the other discrete symmetries it is challenging to find a
mechanism to generate neutrino masses. Finally, the setup also satisfies the
constraints to exhibit all of the $\mathbf{Z}_6$ symmetries, i.e.
\begin{align}
L^2_6  R_6   & = 3 U(1)_a + 3 U(1)_b + 19 U(1)_c + 23 U(1)_d\\
L^2_6  R^3_6 & =   U(1)_a +   U(1)_b + 17 U(1)_c +  5 U(1)_d\\
L^2_6  R^5_6 & =   U(1)_a +   U(1)_b +  9 U(1)_c + 13 U(1)_d\\
R_6         & =   U(1)_a +   U(1)_b +    U(1)_c + 21 U(1)_d\,\,,
\end{align}
where $R_6$ originates from $U^{add}(1)$.
In contrast to the solution \# 1 here the proton hexality may be realized as a subgroup of a larger symmetry, namely a combination of the abelian gauge symmetry $U^{add}(1)$ and the discrete symmetry $L_3 R_3 $. In a concrete realization of this setup the $B\wedge F$ couplings may break the $U^{add}(1)$ down to matter parity $R_2$ and thus only Proton hexality survives in the low energy limit. In case a larger symmetry survives the $B\wedge F$ couplings one needs a dynamical mechanism for the larger symmetry to break down to Proton hexality since otherwise the generation of a Weinberg operator and $\mu$-term is not allowed.

The solution \# $3$ of table \ref{table four stack model  1/3a - 1/2 b+1 d} may
exhibit an additional $U^{add}(1)=U(1)_d$ which potentially remains massless,
i.e. it satisfies the constraints \eqref{eq massless constraint abelian}.
However, the presence of such an abelian gauge symmetry would spoil the model,
since it would forbid various desired Yukawa couplings. Even worse there exists
no discrete subgroup of the abelian gauge symmetry $U^{add}(1)$ that would allow
the desired Yukawa couplings. Thus in a concrete realization it must be absent.
The local D-brane configuration however does allow for a discrete $\mathbf{Z}_2$
that allows all desired Yukawa couplings, the matter parity $R_2$, given by
\begin{align}
R_2 = U(1)_a + U(1)_b + U(1)_c + U(1)_d 
\end{align}
which forbids all R-parity violating couplings.

The solution \# 4 of table \ref{table four stack model  1/3a - 1/2 b+1 d} may
exhibit two additional $U(1)$'s given by
\begin{align}
U^{add}_1(1)= U(1)_b - 2 U(1)_c \qquad\text{and}\qquad  U^{add}_2(1) = U(1)_d
\end{align}
where the latter cannot survive as a gauge symmetry since it would forbid all
desired Yukawa couplings. On the other hand the abelian gauge symmetry
$U^{add}_1(1)$ does allow all superpotential terms. The B-L symmetry is given by
$U_{B-L}(1) = \frac{1}{2} U_Y(1) - \frac{1}{4}  U^{add}_1(1)$ in terms of the hypercharge and the additional $U^{add}_1(1)$. One finds for this
configuration that the discrete subgroup of the two abelian gauge symmetries
$U(1)_Y$ and $U^{add}_1(1)$ do give rise to matter parity $R_2$,
to the $\mathbf{Z}_3$ symmetry $R_3$ and to the $\mathbf{Z}_6$ symmetry $R_6$
These discrete gauge symmetries are realized as the following linear
combinations
\begin{align}
R_2& = U(1)_a + U(1)_b + U(1)_c + U(1)_d  \\
R_3& = U(1)_a + U(1)_b + U(1)_c \\
R_6& = U(1)_a + U(1)_b + U(1)_c +9 U(1)_d\,\,,
\end{align}
While $R_2$ forbids all R-parity violating couplings in this local D-brane
configuration the absence of dimension 5 proton decay operators is rather
accidental and does not originate from a discrete gauge symmetry.

\subsubsection{Hypercharge $U(1)_Y= \frac{1}{6} U(1)_a +\frac{1}{2} U(1)_c -\frac{1}{2}U(1)_d $}

\begin{table}[h]
\scalebox{.65}{
\begin{tabular}{|c|c|c|c|c|c|c|c|c|c|c|c|c|c|c|c|c|c|c|c|c|c|}\hline
\multirow{2}{*}{\hspace{-1mm}$\#$}\hspace{-1mm} & \multicolumn{1}{|c}{$Q_L$} & \multicolumn{2}{|c}{$D_R$} & \multicolumn{2}{|c}{$U_R$} & \multicolumn{2}{|c}{$L$} & \multicolumn{3}{|c}{$E_R$} & \multicolumn{3}{|c}{$N_R$} & \multicolumn{2}{|c}{$H_u$} & \multicolumn{2}{|c|}{$H_d$}\\
\cline{2-18}
& \hspace{-1.5mm}$(\fund_a,\ov\fund_{b})$\hspace{-1.5mm} 
& \hspace{-1.5mm}$(\ov\fund_{a},\fund_c)$\hspace{-1.5mm} & \hspace{-1.5mm}$(\ov\fund_{a},\ov \fund_d)$\hspace{-1.5mm} 
& \hspace{-1.5mm}$(\ov\fund_{a},\ov\fund_{c})$\hspace{-1.5mm} & \hspace{-1.5mm}$(\ov\fund_{a}, \fund_d)$\hspace{-1.5mm} 
& \hspace{-1.5mm}$(\fund_b,\ov\fund_{c})$\hspace{-1.5mm} & \hspace{-1.5mm}$(\fund_b, \fund_d)$\hspace{-1.5mm} 
& \hspace{-1.5mm}$(\fund_c,\ov \fund_d)$\hspace{-1.5mm} & \hspace{-1.5mm}${\Ysymm}_c $\hspace{-1.5mm} & \hspace{-1.5mm}$\ov \Ysymm_d$\hspace{-1.5mm} 
& $\ov{\Yasymm}_b$ $\phantom{I^{I^{I^I}}}$\hspace{-9mm} & \hspace{-1.5mm}$(\fund_c, \fund_d)$\hspace{-1.5mm} & \hspace{-1.5mm}$(\ov \fund_{c}, \ov \fund_d)$\hspace{-1.5mm} 
& \hspace{-1.5mm}$(\fund_b, \fund_c)$\hspace{-1.5mm} & \hspace{-1.5mm}$(\fund_b,\ov \fund_d)$\hspace{-1.5mm} 
& \hspace{-1.5mm}$(\fund_b,\ov\fund_{c})$\hspace{-1.5mm} & \hspace{-1.5mm}$(\ov\fund_{b},\ov\fund_{c})$\hspace{-1.5mm} 
\\
\hline\hline
1&3&3&0&0&3&0&3&0&0&3&2&0&1&0&1&1&0\\ \hline
2&3&3&0&0&3&0&3&1&0&2&2&1&0&0&1&1&0\\ \hline
3&3&3&0&2&1&0&3&2&1&0&2&1&0&0&1&1&0\\ \hline
4&3&3&0&2&1&0&3&0&2&1&2&1&0&0&1&1&0\\ \hline
5&3&3&0&3&0&0&3&2&1&0&2&0&1&1&0&1&0\\ \hline
6&3&3&0&3&0&0&3&0&2&1&2&0&1&1&0&1&0\\ \hline
7&3&3&0&3&0&0&3&1&2&0&2&1&0&1&0&1&0\\ \hline
8&3&0&3&0&3&3&0&2&0&1&3&0&0&0&1&0&1\\ \hline
9&3&0&3&0&3&3&0&0&1&2&3&0&0&0&1&0&1\\ \hline
10&3&0&3&1&2&3&0&3&0&0&3&0&0&1&0&0&1\\ \hline
11&3&0&3&1&2&3&0&1&1&1&3&0&0&1&0&0&1\\ \hline
12&3&0&3&3&0&3&0&0&3&0&3&0&0&1&0&0&1\\ \hline
\end{tabular}}
\caption{{\small MSSM + 3 $N_R$ spectrum for setups with $U(1)_Y= \frac{1}{6} U(1)_a +\frac{1}{2} U(1)_c -\frac{1}{2}U(1)_d$.}
\label{table MSSM madrid embedding}
}
\end{table}

All solutions of table \ref{table MSSM madrid embedding} apart from \# 3, \# 4, \# 10 and
\# 11 exhibit the discrete $\mathbf{Z}_3$ symmetry $L_3 R_3$, i.e. Baryon
triality. For all these solutions the Baryon triality is given by the same
linear combination, namely
\begin{align}
L_3 R_3 = 2 U(1)_c +4 U(1)_d\,\,.
\end{align}
Apart from those discrete symmetry $L_3 R_3$ only the solutions \# 1 and \# 12 may
contain more discrete symmetries. They both may exhibit an additional massless
$U^{add}(1)$\footnote{For solution \# 1 the additional $U(1)$ takes the form
$U^{add}(1)=-U(1)_b-2U(1)_d$ while for solution \# 12 it is given by
$U^{add}(1)=U(1)_b-2U(1)_c$. The form of the B-L symmetry is given by  $U_{B-L}(1) = -U_Y (1)+ \frac{1}{2} U^{add}(1)$ and $U_{B-L}(1) = -U_Y (1)- \frac{1}{2} U^{add}(1)$, respectively.} which allows all desired Yukawa couplings.
Moreover, it turns out that both solutions do give rise to matter parity $R_2$
and all possible $\mathbf{Z}_3$ and $\mathbf{Z}_6$ symmetries. 

For solution \# 1 of table \ref{table MSSM madrid embedding} the matter parity
takes the form
\begin{align}
R_2 = U(1)_a + U(1)_b + 3 U(1)_c + 7 U(1)_d 
\end{align}
while the $\mathbf{Z}_3$ symmetries are given by
\begin{align}
R_3      & = U(1)_a + U(1)_b + 3 U(1)_c + 5 U(1)_d    \\
L_3      & = U(1)_a + U(1)_b + U(1)_c + U(1)_d     \\
L_3 R_3  & = 2 U(1)_c+ 4 U(1)_d \\
L_3 R^2_3& = U(1)_a + U(1)_b +  5 U(1)_c + 3 U(1)_d \,\,.
\end{align}
The $\mathbf{Z}_6$ symmetries read
\begin{align}
R_6         &=  U(1)_a +   U(1)_b +  3 U(1)_c + 23 U(1)_d \\
L^2_6 R_6   &=  U(1)_a +   U(1)_b + 11 U(1)_c + 15 U(1)_d \\
L^2_6 R^3_6 &=  U(1)_a +   U(1)_b + 19 U(1)_c +  7 U(1)_d \\
L^2_6 R^5_6 &= 3U(1)_a + 3 U(1)_b +   U(1)_c +   5 U(1)_d \,\,.
\end{align}

For solution \# 12 of table \ref{table MSSM madrid embedding} the matter parity is
given by the linear combination
\begin{align}
R_2 =  U(1)_a + U(1)_b +  U(1)_c + 5 U(1)_d 
\end{align}
The $\mathbf{Z}_3$ symmetries take the form
\begin{align}
R_3       & = U(1)_a + U(1)_b + U(1)_c+3 U(1)_d    \\
L_3       & = U(1)_a + U(1)_b + 5 U(1)_c+5 U(1)_d     \\
L_3 R_3   & = 2 U(1)_c+ 4U(1)_d \\
L_3 R^2_3 & = U(1)_a + U(1)_b + 3U(1)_c+ U(1)_d \,\,.
\end{align}
and the $\mathbf{Z}_6$ symmetries read 
\begin{align}
R_6         & =   U(1)_a +   U(1)_b +    U(1)_c + 21 U(1)_d \\
L^2_6 R_6   & =   U(1)_a +   U(1)_b +  9 U(1)_c + 13 U(1)_d \\
L^2_6 R^3_6 & =   U(1)_a +   U(1)_b + 17 U(1)_c +  5 U(1)_d \\
L^2_6 R^5_6 & = 3 U(1)_a + 3 U(1)_b + 11 U(1)_c +  7 U(1)_d \,\,.
\end{align}

Beyond the discrete $\mathbf{Z}_2$, $\mathbf{Z}_3$, $\mathbf{Z}_6$ gauge
symmetries as well as the subgroups of the additional $U^{add}(1)$ both
solutions, \# 1 and \# 12, do not possess any further family dependent discrete gauge symmetries.

Again in contrast to the solution \# 1 of table \ref{table four stack model  1/3a - 1/2 b+1 d} the Proton hexality in solution \# 1 and \# 12 may be realized as a subgroup of a larger symmetry, namely a combination of the abelian gauge symmetry $U^{add}(1)$ and the discrete symmetry, matter parity $ R_2$. In a concrete realization of this setup the $B\wedge F$ couplings may break the $U^{add}(1)$ down to matter parity $R_2$ and thus only Proton hexality survives in the low energy limit. However, in case of a larger symmetry surviving the Green-Schwarz mechanism one needs a dynamical mechanism for the larger symmetry to break down to Proton hexality. Otherwise one would face severe difficulties in the generation of a Weinberg operator and $\mu$-term, since the larger symmetry does forbid them.

\subsubsection{$SU(2)$ realized as $Sp(2)$ with $U(1)_Y= \frac{1}{6} U(1)_a +\frac{1}{2} U(1)_c -\frac{1}{2}U(1)_d$}

Since $Sp(2)$ is isomorphic to $SU(2)$ we can realize the $SU(2)$ of the MSSM as a  D-brane stack wrapping 
an orientifold invariant cycle. A D-brane stack that wraps an orientifold invariant cycle, thus giving rise 
to a $Sp(2)$ symmetry does not contain a $U(1)$ gauge factor that can contribute to the hypercharge or a 
discrete symmetry.  Moreover, it should be noted that the tadpole constraint for the $Sp(2)$ stack does 
not give any constraints on the transformation behaviour of the chiral matter fields. While the last 
statement seems to suggest that one finds multiple local D-brane configurations that satisfy the severe 
top-down constraints and exhibiting a (semi-) realistic phenomenology the systematic search performed in 
\cite{Cvetic:2009yh} finds only one configuration displayed in table \ref{table spectrum four stack model Sp2 1/6a+1/2c-1/2d}.

\begin{table}[h]
\center
\hspace{0.00cm}
\scalebox{0.85}{
\begin{tabular}{|c|c|c|c|c|c|c|c|c|}\hline
\multirow{2}{*}{$\#$}&$Q_L$&$D_R$&$U_R$&$L$&$E_R$&$N_R$&$H_u$&$H_d$\\
\cline{2-9}
& $(\fund_a,\fund_b)$ 
& $(\ov\fund_{a},\fund_c)$ 
& $(\ov\fund_{a},\ov\fund_{c})$ 
& $(\fund_b,\fund_d)$ 
& $(\fund_c,\ov\fund_{d})$ 
& $(\ov\fund_{c},\ov\fund_{d})$ 
& $(\fund_b,\fund_c)$ 
& $(\fund_b,\ov \fund_c)$ 
\\
\hline
\hline
1 &3&3&3&3&3&3&1&1\\ \hline		
\end{tabular}}
\caption{\small MSSM + 3 $N_R$ spectrum for setups with $U(1)_Y=\frac{1}{6}\,U(1)_a+\frac{1}{2}\,U(1)_c-\frac{1}{2} U(1)_d$.
\label{table spectrum four stack model Sp2 1/6a+1/2c-1/2d}}
\end{table}

This model may exhibit a massless $U^{add}(1)$  which is given by
\begin{align}
U^{add}(1)= U(1)_c\,\,.
\end{align}
and contains the discrete symmetries $R_2$, $R_3$ and $R_6$. The abelian gauge
symmetry $U^{add}(1)$ can be combined with $U(1)_Y$ such that it gives the $B-L$
symmetry
\begin{align}
U(1)_{B-L} = 2 U(1)_Y - U^{add}(1)\,\,.
\end{align}
In addition the configuration displayed in table \ref{table spectrum four stack
model Sp2 1/6a+1/2c-1/2d}
 may exhibit the $\mathbf{Z}_3$ symmetries  $L_3$, $L_3 R_3$ and $L_3 R^2_3$
given by the linear combination
\begin{align}
L_3       & = 2 U(1)_d\\
L_3 R_3   & = 2 U(1)_c +4 U(1)_d\\
L_3 R^2_3 & =   U(1)_c +4 U(1)_d
\end{align}
as well as the $\mathbf{Z}_6$ discrete symmetries $L^2_6 R_6$, 
$L^2_6 R^3_6$ and $L^2_6 R^5_6$ which take the form
\begin{align}
L^2_6 R_6   & =   U(1)_c +4 U(1)_d\\
L^2_6 R^3_6 & = 3 U(1)_c +4 U(1)_d\\
L^2_6 R^5_6 & =  U(1)_c +8 U(1)_d\,\,.
\end{align} 
As before Proton hexality may appear as a subgroup of a larger symmetry depending on the details of the St{\"u}ckelberg-type couplings in a concrete realization. In that case one needs a dynamical mechanism to further break the larger symmetry to Proton hexality in order to allow for a $\mu$-term and a Weinberg operator.  

Beyond those discrete gauge symmetries the local D-brane configuration does not
exhibit any additional discrete gauge symmetry. In particular the D-brane setup
does not possess any family dependent discrete gauge
symmetries.

\subsection{Summary of the results \label{sec summary}}

Let us give a brief summary of the results of the systematic bottom-up search
performed above. The first thing to note is that we do not find in any of the
intriguing four stack quivers family dependent discrete gauge symmetries that
allow for the desired Yukawa couplings $Q_L H_u U_R$, $Q_L H_d D_R$ and $L H_d
E_R$. This is somewhat not expected since specifically the leptons in those
D-brane configurations do arise from different intersections of D-brane stacks,
and thus transform differently under the anomalous $U(1)$ factors. Nevertheless
after determining the discrete charge of all matter fields all generations do
have the same charge, even though their D-brane origin is significantly different.

The second observation is that we do not find any discrete $\mathbf{Z}_9$ and
$\mathbf{Z}_{18}$ symmetries for the local MSSM D-brane configurations, which can appear in the pure field theoretical
approach. This is due to the more constraining conditions for the appearance of
discrete symmetries in D-brane compactifications. 
\begin{table}[h]
\begin{center}
\scalebox{.80}{
\begin{tabular}{|c|c|c|c||c||c|c|c|c||c|c|c|c|}
\hline
Spectrum                         &Hypercharge                                                    & \hspace{-1.5mm}Table\hspace{-1.5mm}                         &\hspace{-1mm}$\#$\hspace{-1mm} & $R_2$ & $L_3 R_3 $ & $R_3 $ & $L_3$ & $L_3 R^2_3 $& $L^2_6 R^5_6 $ & $R_6 $ & $L^2_6 R^3_6 $ & $L^2_6 R_6 $\\ \hline
\hline
\multirow{4}{*}{MSSM}            &\multirow{3}{*}{$\left(-\frac{1}{3},-\frac{1}{2},0,1\right)$}  &\multirow{3}{*}{\ref{four stack two}}                        &\multirow{1}{*}{1} & &$\checkmark$ & & & & & & &  \\ 
\cline{4-13}
                                 &                                                               &                                                             & 2                 & &$\checkmark$ & & & & & & & \\ 
\cline{4-13}
                                 &                                                               &                                                             & 3                 & & & & & & & & &\\ \cline{2-13} \\[-1em]\cline{2-13}
                                 &$\left(-\frac{1}{3},-\frac{1}{2},0, 1\right)$                  &\ref{four stack three}                                       & 1                 & & & & & & & & &\\ \hline \hline
\multirow{33}{*}{MSSM + 3 $N_R$} &$\left(\frac{1}{6}, 0,\frac{1}{2},-\frac{3}{2}\right)$         &\ref{table spectrum four stack model 1/6a+1/2c-3/2d}         & 1                 &$\checkmark$ &  & $\checkmark$& & & & $\checkmark$ & & \\ 
\cline{2-13} \\ [-1em] \cline{2-13}
                                 &\multirow{12}{*}{$\left(-\frac{1}{3},-\frac{1}{2},0, 0\right)$}&\multirow{12}{*}{\ref{table four stack model  1/3a - 1/2 c}} & 1                 &$\checkmark$ & & & & & & & &\\
\cline{4-13}
                                 &                                                               &                                                             & 2                 &$\checkmark$ & &$\checkmark$ & & & &$\checkmark$ & &\\
\cline{4-13}
                                 &                                                               &                                                             & 3                 &$\checkmark$& &  & & & &  & & \\
\cline{4-13}
                                 &                                                               &                                                             & 4                 & & & & & & & & &\\
\cline{4-13}
                                 &                                                               &                                                             & 5                 & & & & & & & & &\\
\cline{4-13}
                                 &                                                               &                                                             & 6                 &$\checkmark$ & & & & & & & & \\
\cline{4-13}
                                 &                                                               &                                                             & 7                 &$\checkmark$ & & & & & & & & \\
\cline{4-13}
                                 &                                                               &                                                             & 8                 & & & & & & & & & \\
\cline{4-13}
                                 &                                                               &                                                             & 9                 &$\checkmark$ & & & & & & & &\\
\cline{4-13}
                                 &                                                               &                                                             & 10                & & & & & & & & & \\
\cline{4-13}
                                 &                                                               &                                                             & 11                &$\checkmark$ & & & & & & & & \\
\cline{4-13}
                                 &                                                               &                                                             & 12                &$\checkmark$ & & & & & & & & \\\cline{2-13}
\\[-1em]\cline{2-13}
                                 &\multirow{4}{*}{$\left(-\frac{1}{3},-\frac{1}{2},0, 1\right)$} &\multirow{4}{*}{\ref{table four stack model  1/3a - 1/2 b+1 d}} & 1 &$\checkmark$ & $\checkmark$ & & & & $\checkmark$ & & &\\
\cline{4-13}
                                 &                                                               &                                                                & 2 &$\checkmark$ & $\checkmark$ & $\checkmark$ & $\checkmark$ & $\checkmark$ & $\checkmark$ & $\checkmark$ & $\checkmark$ & $\checkmark$ \\
\cline{4-13}& & &3  & $\checkmark$& & & & & & & &\\
\cline{4-13}& & &4  & $\checkmark$ & & $\checkmark$& & & & $\checkmark$ & &\\
\cline{2-13} \\[-1em]\cline{2-13}
&\multirow{12}{*}{$\left(\frac{1}{6},0,\frac{1}{2},-\frac{1}{2}\right)$} & \multirow{12}{*}{\ref{table MSSM madrid embedding}} & 1  &$\checkmark$&$\checkmark$ & $\checkmark$& $\checkmark$& $\checkmark$&$\checkmark$ & $\checkmark$& $\checkmark$& $\checkmark$\\
\cline{4-13}&& & 2  & & $\checkmark$ & & & & & & &\\
\cline{4-13}& && 3  & &  & & & & & & & \\
\cline{4-13}& & &4  & &  & & & & & & &\\
\cline{4-13}& & &5  & & $\checkmark$ & & & & & & &\\
\cline{4-13}& & &6  & & $\checkmark$ & & & & & & &\\
\cline{4-13}& & &7   & & $\checkmark$ & & & & & & &\\
\cline{4-13}& & &8  & &  $\checkmark$& & & & & & & \\
\cline{4-13}& & &9   & &  $\checkmark$ & & & & & & &\\
\cline{4-13}& & &10 & & & & & & & & & \\
\cline{4-13}& & &11 & &   & & & & & & & \\
\cline{4-13}& & &12  &$\checkmark$ &$\checkmark$ & $\checkmark$& $\checkmark$&$\checkmark$&$\checkmark$ & $\checkmark$& $\checkmark$& $\checkmark$\\
\cline{2-13} \\[-1em]\cline{2-13}
& $\left(\frac{1}{6},0,\frac{1}{2},-\frac{1}{2}\right)$ & \ref{table spectrum four stack model Sp2 1/6a+1/2c-1/2d} & 1  &  $\checkmark$&  $\checkmark$&  $\checkmark$& $\checkmark$ & $\checkmark$ &$\checkmark$  & $\checkmark$ &  $\checkmark$ &$\checkmark$ \\ \hline
\end{tabular}}
\caption{{\small The table summarizes our findings on the search of discrete
gauge symmetries in promising local D-brane setups. The symbol $\checkmark$
denotes the potential presence of a discrete gauge symmetry for the respective
local D-brane setup. Matter parity is given by $R_2$, Baryon triality by $L_3 R_3$ and Proton
hexality by $L^2_6 R^5_6$.\label{table summary}}}
\end{center}
\end{table}

Table \ref{table summary} displays for each quiver the potential appearing
discrete symmetries. It shows that matter parity $R_2$ is favoured for the
hypercharge embeddings 
\begin{align}
U(1)_Y & = -\frac{1}{3} U(1)_a -\frac{1}{2} U(1)_b \qquad \text{and} \\
U(1)_Y & = -\frac{1}{3} U(1)_a -\frac{1}{2} U(1)_b+U(1)_d
\label{eq matter hypercharge}
\end{align}
which appears for almost all D-brane setups with these hypercharge embeddings.
For the hypercharge embedding in eq.~\eqref{eq matter hypercharge} $U(1)_Y =
-\frac{1}{3} U(1)_a -\frac{1}{2} U(1)_b$ there is only one configuration out of
12 that allows for a $\mathbf{Z}_3$ and $\mathbf{Z}_6$ discrete symmetry. On the
other hand for the hypercharge embedding $U(1)_Y = -\frac{1}{3} U(1)_a
-\frac{1}{2} U(1)_b+U(1)_d$ we do find for each realization a $\mathbf{Z}_3$
symmetry, but only in two cases it is Baryon triality. Those local D-brane
configurations also allow  for Proton hexality.

For the Madrid embedding
\begin{align*}
U(1)_Y = \frac{1}{6} U(1)_a +\frac{1}{2} U(1)_c-\frac{1}{2} U(1)_d \,\,.
\end{align*}
almost all realizations have the potential to exhibit Baryon triality. However,
the presence of matter parity is highly suppressed. Only for two setups we
also find matter parity realized. Hence, those quivers pass the constraints to
exhibit Proton hexality.

Summarizing we find only five setups that have the potential to exhibit Proton
hexality, which is a particular intriguing discrete symmetry since it forbids
all R-parity violating terms as well as all dangerous dimension 5 proton decay
operators. This suggests that the presence of Proton hexality in D-brane
compactifications is rather suppressed.

Finally, one observes a similar pattern as in the field theoretical approach,
namely that the presence of discrete $\mathbf{Z}_6$ symmetries is tied to the
presence of $\mathbf{Z}_2$ and $\mathbf{Z}_3$ symmetries. We find the same
relations as in pure field theory
\begin{align}
R_2 \times    L_3  R_3     &\cong  L^2_6 R^5_6\\
R_2 \times     R_3        &\cong R_6  \\
R_2 \times     L_3        &\cong  L^2_6 R^3_6\\
R_2 \times      L_3R^2_3   &\cong  L^2_6 R_6\,\,.
\end{align} 
Thus, the presence of $R_2$ along with a discrete $\mathbf{Z}_3$ symmetry implies
the presence of a $\mathbf{Z}_6$ symmetry.

\section{Conclusions \label{sec conclusion}}
We study the presence of discrete gauge symmetries in D-brane compactifications.
We translate the conditions for the presence of a discrete gauge symmetry in
D-brane compactifications laid out in \cite{BerasaluceGonzalez:2011wy} into
constraints on the transformation behaviour of the chiral matter fields. This
allows for a bottom-up search, a search that does not require the knowledge of
any features of the compactification manifold, for local D-brane configurations
with respect to discrete gauge symmetries. 

After establishing those constraints on the transformation behaviour of the
chiral matter fields we perform a systematic search for discrete gauge
symmetries within a class of promising local D-brane quivers based on four stacks
of D-branes. Those local configurations, that are consistent with the global consistency conditions, were found in
\cite{Cvetic:2009yh,Cvetic:2009ng,Cvetic:2010mm} and exhibit the exact MSSM spectrum or the exact MSSM spectrum plus three right-handed neutrinos. Within this
class of intriguing four stack quivers there is no quiver that allows for a
family dependent discrete gauge symmetry. Moreover, none of the local MSSM D-brane
configurations exhibits a discrete $\mathbf{Z}_9$ and $\mathbf{Z}_{18}$ gauge
symmetry, which, on the other hand, were found in \cite{Dreiner:2005rd} using a
pure field theoretical approach. This confirms one of our earlier findings that
the constraints on the transformation behaviour of the chiral matter fields for
having a discrete gauge symmetry in D-brane compactifications goes beyond the
four-dimensional discrete gauge anomaly conditions used in 
\cite{Dreiner:2005rd}.

Our search reveals that all $\mathbf{Z}_2$, $\mathbf{Z}_3$ and $\mathbf{Z}_6$
discrete gauge symmetries found in \cite{Dreiner:2005rd} can be also realized in
the local D-brane configurations. We find that the realization of discrete
symmetries depends on the hypercharge embedding of the D-brane
configuration. For instance while the Madrid embedding favours Baryon triality
it disfavours matter parity. The presence of Proton hexality, i.e. the
simultaneous presence of matter parity and Baryon triality, is rather suppressed
and only realized for five of the intriguing four D-brane-stack quivers. In
those quivers the absence of R-parity and disastrous dimension 5 proton decay
operators is not accidental, but can be explained by the presence of a discrete
gauge symmetry. 

It would be interesting to extend this analysis to local semi-realistic D-brane configurations with more than 4 D-brane stacks. Specifically, it would be interesting to see whether one can find family dependent discrete gauge symmetries in those realizations. Furthermore, another intriguing avenue is to extend the analysis to the NMSSM \cite{Cvetic:2010dz} and GUT realizations of the MSSM \cite{Anastasopoulos:2010hu} as well as extending it to local D-brane configurations with additional exotics \cite{Cvetic:2011iq}.

Finally, we would like to comment on the limits of the bottom-up approach
applied here. The discrete gauge symmetries considered here purely originate
from the anomalous $U(1)$ factors carried by each D-brane stack. In addition
there may be abelian or even non-abelian gauge factors arising from isometries of
the compactification manifold which can lead to abelian and non-abelian discrete
gauge symmetries in the low energy effective action
\cite{BerasaluceGonzalez:2012vb}. The consideration of discrete symmetries
originating from isometries, requires the specification of the properties of the
compactification and thus goes beyond the scope of this work.

\section*{Acknowledgements}
We acknowledge J. Halverson, G. Honecker, L.E. Ib{\'a}{\~n}ez, E. Kiritsis, P. Langacker, B. Schellekens for interesting discussions and correspondence. 
P.~A. is supported by the Austrian Science Fund (FWF) program M 1428-N27. 
The work of M.C. is supported in part by DOE grant DE-SC0007901, the Fay
R. and Eugene L. Langberg Endowed Chair and the Slovenian Research
Agency (ARRS).
The work of P.~V and R.~R. was partly supported by the German Science Foundation (DFG) under the Collaborative Research Center (SFB) 676 ``Particles, Strings and the Early Universe".
P.~A., M.~C. and R.~R. are grateful to the organizers of the ``String Phenomenology TH institute''
at CERN for hospitality during parts of this work. R.~R. and P.~V. thank the organizers of the 4th Bethe Center Workshop on ``Unification and String Theory" in Bad Honnef for hospitality during last stage of this work.
R.~R. is also grateful to ``Tor Vergata" for hospitality during the last part of this work.

\newpage

\appendix

\section{Bottom-up constraints
\label{app bottom up constraints}}
In this appendix we display all the bottom-up constraints which were imposed in
the search of realistic local D-brane configurations. Apart from the top-down
constraints \eqref{eq constraint1}, \eqref{eq constraint2}  and \eqref{eq
massless constraint abelian} that the spectrum has to satisfy we furthermore
require a few phenomenological constraints to be satisfied

\begin{itemize}
\item[$\bullet$] The MSSM superpotential couplings 
\begin{align}  
Q_L\, H_u \,U_{R} \qquad  Q_L\, H_d \,D_{R} \qquad L \, H_d \, E_R 
\end{align}
are either realized perturbatively or in case they violate global $U(1)$
selection rules and thus are perturbatively forbidden they will be induced by
D-instantons, such that all three families of quarks and charged leptons acquire
masses. 

\item[$\bullet$] We require that the D-brane quiver exhibits a mechanism which
accounts for the neutrino masses
\cite{Blumenhagen:2006xt,Ibanez:2006da,Cvetic:2007ku,Ibanez:2007rs,
Antusch:2007jd,Cvetic:2007qj,Cvetic:2008hi}.

\item[$\bullet$] We demand the absence of the R-parity violating couplings
\begin{equation}
D_R\, D_R\, U_R \qquad  L \, L\, E_R \qquad  Q_L\, L\, D_R \qquad  L\, H_u
\end{equation}
on the perturbative level and furthermore, require that they are not generated
by a D-instanton whose presence is required to generate a perturbatively
forbidden, but desired, MSSM superpotential couplings. 

\item[$\bullet$] We demand that none of the D-instantons required to generate
desired Yukawa couplings does induce a tadpole $N_R$.

\item[$\bullet$]
We rule out setups which lead to a large family mixing in the quark Yukawa
couplings \cite{Ibanez:2008my,Leontaris:2009ci,Cvetic:2009yh,Anastasopoulos:2009mr,Cvetic:2009ez}.

\item[$\bullet$] We demand the absence of the dangerous dimension $5$ proton
decay operators
\begin{equation}
U_R\,U_R\,D_R\,E_R \qquad  \text{and} \qquad Q_L\,Q_L\, Q_L\, L
\end{equation}
on the perturbative level and additionally require that they are not generated
by a D-instanton whose presence is required to generate a perturbatively
forbidden, but desired, MSSM superpotential couplings.

\end{itemize}

\clearpage \nocite{*}

\bibliographystyle{JHEP}
\bibliography{discretref.bib}

\providecommand{\href}[2]{#2}\begingroup\raggedright\begin{thebibliography}{10}

\bibitem{Banks:1988yz}
T.~Banks and L.~J. Dixon, {\it {Constraints on String Vacua with Space-Time
  Supersymmetry}},  {\em Nucl.Phys.} {\bf B307} (1988) 93--108.

\bibitem{Abbott:1989jw}
L.~Abbott and M.~B. Wise, {\it {WORMHOLES AND GLOBAL SYMMETRIES}},  {\em
  Nucl.Phys.} {\bf B325} (1989) 687.

\bibitem{Coleman:1989zu}
S.~R. Coleman and K.-M. Lee, {\it {WORMHOLES MADE WITHOUT MASSLESS MATTER
  FIELDS}},  {\em Nucl.Phys.} {\bf B329} (1990) 387.

\bibitem{Kallosh:1995hi}
R.~Kallosh, A.~D. Linde, D.~A. Linde, and L.~Susskind, {\it {Gravity and global
  symmetries}},  {\em Phys.Rev.} {\bf D52} (1995) 912--935,
  [\href{http://xxx.lanl.gov/abs/hep-th/9502069}{{\tt hep-th/9502069}}].

\bibitem{Banks:2010zn}
T.~Banks and N.~Seiberg, {\it {Symmetries and Strings in Field Theory and
  Gravity}},  {\em Phys.Rev.} {\bf D83} (2011) 084019,
  [\href{http://xxx.lanl.gov/abs/1011.5120}{{\tt arXiv:1011.5120}}].

\bibitem{BerasaluceGonzalez:2011wy}
M.~Berasaluce-Gonz{\'a}lez, L.~E. Ib{\'a}{\~n}ez, P.~Soler, and A.~M. Uranga,
  {\it {Discrete gauge symmetries in D-brane models}},  {\em JHEP} {\bf 1112}
  (2011) 113, [\href{http://xxx.lanl.gov/abs/1106.4169}{{\tt
  arXiv:1106.4169}}].

\bibitem{Green:1984ed}
M.~B. Green and J.~H. Schwarz, {\it {Infinity Cancellations in SO(32)
  Superstring Theory}},  {\em Phys.Lett.} {\bf B151} (1985) 21--25.

\bibitem{Sagnotti:1992qw}
A.~Sagnotti, {\it {A Note on the Green-Schwarz mechanism in open string
  theories}},  {\em Phys.Lett.} {\bf B294} (1992) 196--203,
  [\href{http://xxx.lanl.gov/abs/hep-th/9210127}{{\tt hep-th/9210127}}].

\bibitem{Ibanez:1998qp}
L.~E. Ib{\'a}{\~n}ez, R.~Rabadan, and A.~Uranga, {\it {Anomalous U(1)'s in type
  I and type IIB D = 4, N=1 string vacua}},  {\em Nucl.Phys.} {\bf B542} (1999)
  112--138, [\href{http://xxx.lanl.gov/abs/hep-th/9808139}{{\tt
  hep-th/9808139}}].

\bibitem{Bianchi:2000de}
M.~Bianchi and J.~F. Morales, {\it {Anomalies \& tadpoles}},  {\em JHEP} {\bf
  0003} (2000) 030, [\href{http://xxx.lanl.gov/abs/hep-th/0002149}{{\tt
  hep-th/0002149}}].

\bibitem{Cvetic:2001nr}
M.~Cveti{\v c}, G.~Shiu, and A.~M. Uranga, {\it {Chiral four-dimensional N=1
  supersymmetric type 2A orientifolds from intersecting D6 branes}},  {\em
  Nucl.Phys.} {\bf B615} (2001) 3--32,
  [\href{http://xxx.lanl.gov/abs/hep-th/0107166}{{\tt hep-th/0107166}}].

\bibitem{Antoniadis:2002cs}
I.~Antoniadis, E.~Kiritsis, and J.~Rizos, {\it {Anomalous U(1)s in type 1
  superstring vacua}},  {\em Nucl.Phys.} {\bf B637} (2002) 92--118,
  [\href{http://xxx.lanl.gov/abs/hep-th/0204153}{{\tt hep-th/0204153}}].

\bibitem{Anastasopoulos:2003aj}
P.~Anastasopoulos, {\it {4-D anomalous U(1)'s, their masses and their relation
  to 6-D anomalies}},  {\em JHEP} {\bf 0308} (2003) 005,
  [\href{http://xxx.lanl.gov/abs/hep-th/0306042}{{\tt hep-th/0306042}}].

\bibitem{Anastasopoulos:2004ga}
P.~Anastasopoulos, {\it {Anomalous U(1)s masses in nonsupersymmetric open
  string vacua}},  {\em Phys.Lett.} {\bf B588} (2004) 119--126,
  [\href{http://xxx.lanl.gov/abs/hep-th/0402105}{{\tt hep-th/0402105}}].

\bibitem{Anastasopoulos:2006cz}
P.~Anastasopoulos, M.~Bianchi, E.~Dudas, and E.~Kiritsis, {\it {Anomalies,
  anomalous U(1)'s and generalized Chern-Simons terms}},  {\em JHEP} {\bf 0611}
  (2006) 057, [\href{http://xxx.lanl.gov/abs/hep-th/0605225}{{\tt
  hep-th/0605225}}].

\bibitem{Blumenhagen:2006xt}
R.~Blumenhagen, M.~Cveti{\v c}, and T.~Weigand, {\it {Spacetime instanton
  corrections in 4D string vacua: The Seesaw mechanism for D-Brane models}},
  {\em Nucl.Phys.} {\bf B771} (2007) 113--142,
  [\href{http://xxx.lanl.gov/abs/hep-th/0609191}{{\tt hep-th/0609191}}].

\bibitem{Ibanez:2006da}
L.~Ib{\'a}{\~n}ez and A.~Uranga, {\it {Neutrino Majorana Masses from String
  Theory Instanton Effects}},  {\em JHEP} {\bf 0703} (2007) 052,
  [\href{http://xxx.lanl.gov/abs/hep-th/0609213}{{\tt hep-th/0609213}}].

\bibitem{Blumenhagen:2009qh}
R.~Blumenhagen, M.~Cveti{\v c}, S.~Kachru, and T.~Weigand, {\it {D-Brane
  Instantons in Type II Orientifolds}},  {\em Ann.Rev.Nucl.Part.Sci.} {\bf 59}
  (2009) 269--296, [\href{http://xxx.lanl.gov/abs/0902.3251}{{\tt
  arXiv:0902.3251}}].

\bibitem{Ibanez:2012wg}
L.~Ib{\'a}{\~n}ez, A.~Schellekens, and A.~Uranga, {\it {Discrete Gauge
  Symmetries in Discrete MSSM-like Orientifolds}},  {\em Nucl.Phys.} {\bf B865}
  (2012) 509--540, [\href{http://xxx.lanl.gov/abs/1205.5364}{{\tt
  arXiv:1205.5364}}].

\bibitem{Cvetic:2009yh}
M.~Cveti{\v c}, J.~Halverson, and R.~Richter, {\it {Realistic Yukawa structures
  from orientifold compactifications}},  {\em JHEP} {\bf 12} (2009) 063,
  [\href{http://xxx.lanl.gov/abs/0905.3379}{{\tt arXiv:0905.3379}}].

\bibitem{Cvetic:2009ng}
M.~Cveti{\v c}, J.~Halverson, and R.~Richter, {\it {Mass Hierarchies versus
  proton Decay in MSSM Orientifold Compactifications}},
  \href{http://xxx.lanl.gov/abs/0910.2239}{{\tt arXiv:0910.2239}}.

\bibitem{Cvetic:2010mm}
M.~Cveti{\v c}, J.~Halverson, P.~Langacker, and R.~Richter, {\it {The Weinberg
  Operator and a Lower String Scale in Orientifold Compactifications}},  {\em
  JHEP} {\bf 10} (2010) 094, [\href{http://xxx.lanl.gov/abs/1001.3148}{{\tt
  arXiv:1001.3148}}].

\bibitem{Dreiner:2005rd}
H.~K. Dreiner, C.~Luhn, and M.~Thormeier, {\it {What is the discrete gauge
  symmetry of the MSSM?}},  {\em Phys.Rev.} {\bf D73} (2006) 075007,
  [\href{http://xxx.lanl.gov/abs/hep-ph/0512163}{{\tt hep-ph/0512163}}].

\bibitem{Ibanez:1991pr}
L.~E. Ib{\'a}{\~n}ez and G.~G. Ross, {\it {Discrete gauge symmetries and the
  origin of baryon and lepton number conservation in supersymmetric versions of
  the standard model}},  {\em Nucl.Phys.} {\bf B368} (1992) 3--37.

\bibitem{Ibanez:1991hv}
L.~E. Ib{\'a}{\~n}ez and G.~G. Ross, {\it {Discrete gauge symmetry anomalies}},
   {\em Phys.Lett.} {\bf B260} (1991) 291--295.

\bibitem{Araki:2008ek}
T.~Araki, T.~Kobayashi, J.~Kubo, S.~Ramos-S{\'a}nchez, M.~Ratz, et~al., {\it
  {(Non-)Abelian discrete anomalies}},  {\em Nucl.Phys.} {\bf B805} (2008)
  124--147, [\href{http://xxx.lanl.gov/abs/0805.0207}{{\tt arXiv:0805.0207}}].

\bibitem{Farrar:1978xj}
G.~R. Farrar and P.~Fayet, {\it {Phenomenology of the Production, Decay, and
  Detection of New Hadronic States Associated with Supersymmetry}},  {\em
  Phys.Lett.} {\bf B76} (1978) 575--579.

\bibitem{Lee:2010gv}
H.~M. Lee, S.~Raby, M.~Ratz, G.~G. Ross, R.~Schieren, et~al., {\it {A unique
  $Z_4^R$ symmetry for the MSSM}},  {\em Phys.Lett.} {\bf B694} (2011)
  491--495, [\href{http://xxx.lanl.gov/abs/1009.0905}{{\tt arXiv:1009.0905}}].

\bibitem{Lee:2011dya}
H.~M. Lee, S.~Raby, M.~Ratz, G.~G. Ross, R.~Schieren, et~al., {\it {Discrete R
  symmetries for the MSSM and its singlet extensions}},  {\em Nucl.Phys.} {\bf
  B850} (2011) 1--30, [\href{http://xxx.lanl.gov/abs/1102.3595}{{\tt
  arXiv:1102.3595}}].

\bibitem{Chen:2012jg}
M.-C. Chen, M.~Ratz, C.~Staudt, and P.~K. Vaudrevange, {\it {The mu Term and
  Neutrino Masses}},  {\em Nucl.Phys.} {\bf B866} (2013) 157--176,
  [\href{http://xxx.lanl.gov/abs/1206.5375}{{\tt arXiv:1206.5375}}].

\bibitem{Altarelli:2010gt}
G.~Altarelli and F.~Feruglio, {\it {Discrete Flavor Symmetries and Models of
  Neutrino Mixing}},  {\em Rev.Mod.Phys.} {\bf 82} (2010) 2701--2729,
  [\href{http://xxx.lanl.gov/abs/1002.0211}{{\tt arXiv:1002.0211}}].

\bibitem{Antoniadis:2000ena}
I.~Antoniadis, E.~Kiritsis, and T.~N. Tomaras, {\it {A D-brane alternative to
  unification}},  {\em Phys. Lett.} {\bf B486} (2000) 186--193,
  [\href{http://xxx.lanl.gov/abs/hep-ph/0004214}{{\tt hep-ph/0004214}}].

\bibitem{Aldazabal:2000sa}
G.~Aldazabal, L.~E. Ib{\'a}{\~n}ez, F.~Quevedo, and A.~Uranga, {\it {D-branes
  at singularities: A Bottom up approach to the string embedding of the
  standard model}},  {\em JHEP} {\bf 0008} (2000) 002,
  [\href{http://xxx.lanl.gov/abs/hep-th/0005067}{{\tt hep-th/0005067}}].

\bibitem{Antoniadis:2001np}
I.~Antoniadis, E.~Kiritsis, and T.~Tomaras, {\it {D-brane Standard Model}},
  {\em Fortsch. Phys.} {\bf 49} (2001) 573--580,
  [\href{http://xxx.lanl.gov/abs/hep-th/0111269}{{\tt hep-th/0111269}}].

\bibitem{Ibanez:2008my}
L.~E. Ib{\'a}{\~n}ez and R.~Richter, {\it {Stringy Instantons and Yukawa
  Couplings in MSSM-like Orientifold Models}},  {\em JHEP} {\bf 03} (2009) 090,
  [\href{http://xxx.lanl.gov/abs/0811.1583}{{\tt arXiv:0811.1583}}].

\bibitem{Leontaris:2009ci}
G.~K. Leontaris, {\it {Instanton induced charged fermion and neutrino masses in
  a minimal Standard Model scenario from intersecting D- branes}},  {\em Int.
  J. Mod. Phys.} {\bf A24} (2009) 6035--6049,
  [\href{http://xxx.lanl.gov/abs/0903.3691}{{\tt arXiv:0903.3691}}].

\bibitem{Kiritsis:2009sf}
E.~Kiritsis, M.~Lennek, and B.~Schellekens, {\it {SU(5) orientifolds, Yukawa
  couplings, Stringy Instantons and Proton Decay}},  {\em Nucl.Phys.} {\bf
  B829} (2010) 298--324, [\href{http://xxx.lanl.gov/abs/0909.0271}{{\tt
  arXiv:0909.0271}}].

\bibitem{Anastasopoulos:2010ca}
P.~Anastasopoulos, G.~Leontaris, and N.~Vlachos, {\it {Phenomenological
  Analysis of D-Brane Pati-Salam Vacua}},  {\em JHEP} {\bf 1005} (2010) 011,
  [\href{http://xxx.lanl.gov/abs/1002.2937}{{\tt arXiv:1002.2937}}].

\bibitem{Fucito:2010dk}
F.~Fucito, A.~Lionetto, J.~Morales, and R.~Richter, {\it {Dynamical
  Supersymmetry Breaking in Intersecting Brane Models}},  {\em JHEP} {\bf 1011}
  (2010) 024, [\href{http://xxx.lanl.gov/abs/1007.5449}{{\tt
  arXiv:1007.5449}}].

\bibitem{Cvetic:2010dz}
M.~Cveti{\v c}, J.~Halverson, and P.~Langacker, {\it {Singlet Extensions of the
  MSSM in the Quiver Landscape}},  {\em JHEP} {\bf 1009} (2010) 076,
  [\href{http://xxx.lanl.gov/abs/1006.3341}{{\tt arXiv:1006.3341}}].

\bibitem{Anastasopoulos:2010hu}
P.~Anastasopoulos, G.~Leontaris, R.~Richter, and A.~Schellekens, {\it {SU(5)
  D-brane realizations, Yukawa couplings and proton stability}},  {\em JHEP}
  {\bf 1012} (2010) 011, [\href{http://xxx.lanl.gov/abs/1010.5188}{{\tt
  arXiv:1010.5188}}].

\bibitem{Cvetic:2011iq}
M.~Cveti{\v c}, J.~Halverson, and P.~Langacker, {\it {Implications of String
  Constraints for Exotic Matter and Z' s Beyond the Standard Model}},  {\em
  JHEP} {\bf 1111} (2011) 058, [\href{http://xxx.lanl.gov/abs/1108.5187}{{\tt
  arXiv:1108.5187}}].

\bibitem{Cvetic:2012kv}
M.~Cveti{\v c}, J.~Halverson, and P.~Langacker, {\it {Ultraviolet Completions
  of Axigluon Models and Their Phenomenological Consequences}},
  \href{http://xxx.lanl.gov/abs/1209.2741}{{\tt arXiv:1209.2741}}.

\bibitem{Cvetic:2012kj}
M.~Cveti{\v c}, J.~Halverson, and H.~Piragua, {\it {Stringy Hidden Valleys}},
  \href{http://xxx.lanl.gov/abs/1210.5245}{{\tt arXiv:1210.5245}}.

\bibitem{Forste:2010pf}
S.~F{\"o}rste, H.~P. Nilles, S.~Ramos-S{\'a}nchez, and P.~K. Vaudrevange, {\it
  {Proton Hexality in Local Grand Unification}},  {\em Phys.Lett.} {\bf B693}
  (2010) 386--392, [\href{http://xxx.lanl.gov/abs/1007.3915}{{\tt
  arXiv:1007.3915}}].

\bibitem{Kappl:2010yu}
R.~Kappl, B.~Petersen, S.~Raby, M.~Ratz, R.~Schieren, et~al., {\it
  {String-Derived MSSM Vacua with Residual R Symmetries}},  {\em Nucl.Phys.}
  {\bf B847} (2011) 325--349, [\href{http://xxx.lanl.gov/abs/1012.4574}{{\tt
  arXiv:1012.4574}}].

\bibitem{BerasaluceGonzalez:2012vb}
M.~Berasaluce-Gonz{\'a}lez, P.~C{\'a}mara, F.~Marchesano, D.~Regalado, and
  A.~Uranga, {\it {Non-Abelian discrete gauge symmetries in 4d string models}},
   \href{http://xxx.lanl.gov/abs/1206.2383}{{\tt arXiv:1206.2383}}.

\bibitem{Cvetic:2007ku}
M.~Cveti{\v c}, R.~Richter, and T.~Weigand, {\it {Computation of D-brane
  instanton induced superpotential couplings: Majorana masses from string
  theory}},  {\em Phys.Rev.} {\bf D76} (2007) 086002,
  [\href{http://xxx.lanl.gov/abs/hep-th/0703028}{{\tt hep-th/0703028}}].

\bibitem{Ibanez:2007rs}
L.~Ib{\'a}{\~n}ez, A.~Schellekens, and A.~Uranga, {\it {Instanton Induced
  Neutrino Majorana Masses in CFT Orientifolds with MSSM-like spectra}},  {\em
  JHEP} {\bf 0706} (2007) 011, [\href{http://xxx.lanl.gov/abs/0704.1079}{{\tt
  arXiv:0704.1079}}].

\bibitem{Blumenhagen:2007zk}
R.~Blumenhagen, M.~Cveti{\v c}, D.~L{\"u}st, R.~Richter, and T.~Weigand, {\it
  {Non-perturbative Yukawa Couplings from String Instantons}},  {\em
  Phys.Rev.Lett.} {\bf 100} (2008) 061602,
  [\href{http://xxx.lanl.gov/abs/0707.1871}{{\tt arXiv:0707.1871}}].

\bibitem{Blumenhagen:2005mu}
R.~Blumenhagen, M.~Cveti{\v c}, P.~Langacker, and G.~Shiu, {\it {Toward
  realistic intersecting D-brane models}},  {\em Ann.Rev.Nucl.Part.Sci.} {\bf
  55} (2005) 71--139, [\href{http://xxx.lanl.gov/abs/hep-th/0502005}{{\tt
  hep-th/0502005}}].

\bibitem{Blumenhagen:2006ci}
R.~Blumenhagen, B.~K{\"o}rs, D.~L{\"u}st, and S.~Stieberger, {\it
  {Four-dimensional String Compactifications with D-Branes, Orientifolds and
  Fluxes}},  {\em Phys.Rept.} {\bf 445} (2007) 1--193,
  [\href{http://xxx.lanl.gov/abs/hep-th/0610327}{{\tt hep-th/0610327}}].

\bibitem{Marchesano:2007de}
F.~Marchesano, {\it {Progress in D-brane model building}},  {\em Fortsch.Phys.}
  {\bf 55} (2007) 491--518, [\href{http://xxx.lanl.gov/abs/hep-th/0702094}{{\tt
  hep-th/0702094}}].

\bibitem{Cvetic:2011vz}
M.~Cveti{\v c} and J.~Halverson, {\it {TASI Lectures: Particle Physics from
  Perturbative and Non-perturbative Effects in D-braneworlds}},
  \href{http://xxx.lanl.gov/abs/1101.2907}{{\tt arXiv:1101.2907}}.

\bibitem{Blumenhagen:2000wh}
R.~Blumenhagen, L.~Goerlich, B.~Kors, and D.~Lust, {\it {Noncommutative
  compactifications of type I strings on tori with magnetic background flux}},
  {\em JHEP} {\bf 0010} (2000) 006,
  [\href{http://xxx.lanl.gov/abs/hep-th/0007024}{{\tt hep-th/0007024}}].

\bibitem{Aldazabal:2000dg}
G.~Aldazabal, S.~Franco, L.~E. Ib{\'a}{\~n}ez, R.~Rabad{\'a}n, and A.~Uranga,
  {\it {D = 4 chiral string compactifications from intersecting branes}},  {\em
  J.Math.Phys.} {\bf 42} (2001) 3103--3126,
  [\href{http://xxx.lanl.gov/abs/hep-th/0011073}{{\tt hep-th/0011073}}].

\bibitem{Aldazabal:2000cn}
G.~Aldazabal, S.~Franco, L.~E. Ib{\'a}{\~n}ez, R.~Rabad{\'a}n, and A.~Uranga,
  {\it {Intersecting brane worlds}},  {\em JHEP} {\bf 0102} (2001) 047,
  [\href{http://xxx.lanl.gov/abs/hep-ph/0011132}{{\tt hep-ph/0011132}}].

\bibitem{Blumenhagen:2001te}
R.~Blumenhagen, B.~K{\"o}rs, D.~L{\"u}st, and T.~Ott, {\it {The standard model
  from stable intersecting brane world orbifolds}},  {\em Nucl.Phys.} {\bf
  B616} (2001) 3--33, [\href{http://xxx.lanl.gov/abs/hep-th/0107138}{{\tt
  hep-th/0107138}}].

\bibitem{Cvetic:2001tj}
M.~Cveti{\v c}, G.~Shiu, and A.~M. Uranga, {\it {Three family supersymmetric
  standard - like models from intersecting brane worlds}},  {\em
  Phys.Rev.Lett.} {\bf 87} (2001) 201801,
  [\href{http://xxx.lanl.gov/abs/hep-th/0107143}{{\tt hep-th/0107143}}].

\bibitem{Honecker:2003vq}
G.~Honecker, {\it {Chiral supersymmetric models on an orientifold of Z(4) x
  Z(2) with intersecting D6-branes}},  {\em Nucl.Phys.} {\bf B666} (2003)
  175--196, [\href{http://xxx.lanl.gov/abs/hep-th/0303015}{{\tt
  hep-th/0303015}}].

\bibitem{Honecker:2004kb}
G.~Honecker and T.~Ott, {\it {Getting just the supersymmetric standard model at
  intersecting branes on the Z(6) orientifold}},  {\em Phys.Rev.} {\bf D70}
  (2004) 126010, [\href{http://xxx.lanl.gov/abs/hep-th/0404055}{{\tt
  hep-th/0404055}}].

\bibitem{Gmeiner:2008xq}
F.~Gmeiner and G.~Honecker, {\it {Millions of Standard Models on Z-prime(6)?}},
   {\em JHEP} {\bf 0807} (2008) 052,
  [\href{http://xxx.lanl.gov/abs/0806.3039}{{\tt arXiv:0806.3039}}].

\bibitem{Dijkstra:2004cc}
T.~Dijkstra, L.~Huiszoon, and A.~Schellekens, {\it {Supersymmetric standard
  model spectra from RCFT orientifolds}},  {\em Nucl.Phys.} {\bf B710} (2005)
  3--57, [\href{http://xxx.lanl.gov/abs/hep-th/0411129}{{\tt hep-th/0411129}}].

\bibitem{Anastasopoulos:2006da}
P.~Anastasopoulos, T.~Dijkstra, E.~Kiritsis, and A.~Schellekens, {\it
  {Orientifolds, hypercharge embeddings and the Standard Model}},  {\em
  Nucl.Phys.} {\bf B759} (2006) 83--146,
  [\href{http://xxx.lanl.gov/abs/hep-th/0605226}{{\tt hep-th/0605226}}].

\bibitem{Gmeiner:2005vz}
F.~Gmeiner, R.~Blumenhagen, G.~Honecker, D.~Lust, and T.~Weigand, {\it {One in
  a billion: MSSM-like D-brane statistics}},  {\em JHEP} {\bf 0601} (2006) 004,
  [\href{http://xxx.lanl.gov/abs/hep-th/0510170}{{\tt hep-th/0510170}}].

\bibitem{Antusch:2007jd}
S.~Antusch, L.~E. Ib{\'a}{\~n}ez, and T.~Macri, {\it {Neutrino Masses and
  Mixings from String Theory Instantons}},  {\em JHEP} {\bf 09} (2007) 087,
  [\href{http://xxx.lanl.gov/abs/0706.2132}{{\tt arXiv:0706.2132}}].

\bibitem{Cvetic:2007qj}
M.~Cveti{\v c} and T.~Weigand, {\it {Hierarchies from D-brane instantons in
  globally defined Calabi-Yau Orientifolds}},  {\em Phys. Rev. Lett.} {\bf 100}
  (2008) 251601, [\href{http://xxx.lanl.gov/abs/0711.0209}{{\tt
  arXiv:0711.0209}}].

\bibitem{Cvetic:2008hi}
M.~Cveti{\v c} and P.~Langacker, {\it {D-Instanton Generated Dirac Neutrino
  Masses}},  {\em Phys. Rev.} {\bf D78} (2008) 066012,
  [\href{http://xxx.lanl.gov/abs/0803.2876}{{\tt arXiv:0803.2876}}].

\bibitem{Anastasopoulos:2009mr}
P.~Anastasopoulos, E.~Kiritsis, and A.~Lionetto, {\it {On mass hierarchies in
  orientifold vacua}},  {\em JHEP} {\bf 0908} (2009) 026,
  [\href{http://xxx.lanl.gov/abs/0905.3044}{{\tt arXiv:0905.3044}}].

\bibitem{Cvetic:2009ez}
M.~Cveti{\v c}, J.~Halverson, and R.~Richter, {\it {Mass Hierarchies from MSSM
  Orientifold Compactifications}},  {\em JHEP} {\bf 1007} (2010) 005,
  [\href{http://xxx.lanl.gov/abs/0909.4292}{{\tt arXiv:0909.4292}}].

\end{thebibliography}\endgroup

\end{document}